\documentclass[twocolumn]{aastex3}
\usepackage{graphicx,subfigure}
\usepackage{rotating}
\usepackage{epsfig}
\usepackage{amssymb}
\usepackage{amsmath}
\usepackage[utf8]{inputenc}
\usepackage{array}
\usepackage{ragged2e}
\usepackage{longtable}
\usepackage{supertabular}
\usepackage{subfloat}
\shortauthors{Sriram et al.}

\begin{document}
\title{Constraining the Coronal Heights and Readjustment Velocities based on the Detection of a Few hundred seconds Delays in the Z source GX 17+2 }

\author{K. Sriram}
\affil{Department of Astronomy, Osmania University, Hyderabad 500007, India}
\affil{Korea Astronomy and Space Science Institute, Daejeon 34055, Republic of Korea}
\email{astrosriram@yahoo.co.in}
\author{S. Malu}
\affil{Department of Astronomy, Osmania University, Hyderabad 500007, India}

\author{C. S. Choi}
\affil{Korea Astronomy and Space Science Institute, Daejeon 34055, Republic of Korea}

\begin{abstract}
Neutron star Z type sources provide a unique platform in order to understand the structure of accretion disk-corona geometry emitting close to the Eddington luminosity. Using {\it RXTE} and {\it NuSTAR} satellite data, we performed cross correlation function (CCF) studies in GX 17+2 in order to constrain the size of corona responsible for hard X-rays. From the RXTE data, we found that during horizontal and normal branches, the CCFs show anti-correlated hard (16 - 30 keV) and soft  (2 - 5 keV) X-ray delays of the order of a few tens to hundred seconds with a mean correlation coefficient of 0.42$\pm$0.11. Few observations shows correlated lags and on one occasion coincident with radio emission. We also report an anti-correlated hard X-ray delay of 113$\pm$51 s using the {\it NuSTAR} data of GX 17+2. Based on RXTE data, it was found that soft and hard X-ray fluxes are varying indicating the changes in the disk-corona structure during delays. We bridle the size of corona using relativistic precession, transition layer and boundary layer models. Assuming the delays to be readjustment time scale of disk-corona structure, height of the corona was estimated to be $\sim$ 17--100 km. Assuming that inner region of the truncated disk is occupied by the corona, we constrain the coronal readjustment velocities ({\it v$_{corona}$=$\beta$ v$_{disk}$}, where v$_{disk}$ is the radial velocity component of the disk) of the order of $\beta$=0.06-0.12. This study indicates that the observed delays are primarily dependent on the varying coronal readjustment velocities..  
\end{abstract}
    

\section{Introduction}

The geometry of accretion disk in low-mass X-ray binaries (LMXBs) is well developed viz. truncated accretion disk (Esin et al. 1997; Done et al. 2007) and accretion disk corona models (e.g. Church et al. 2006) but still debatable in terms of location, geometry and especially the variability caused in corona due to the inner accretion flow as well as jet. Though the picture is relatively clear in black hole X-ray binaries (BHBs) but the complexity remains in case of atoll and Z sources (Done et al. 2002; Di Salvo et al. 2002; Church et al. 2006; Church et al. 2014). At low luminosities $\sim$0.001--0.5 L$_{E}$ (Done et al. 2007) atoll sources exhibit two prominent branches i.e. island and banana states in hardness intensity diagram (HID) or color-color diagram (CCD) (Hasinger \& van der Klis 1989; van der Klis 2006). The island state is further divided into upper branch (UB), lower branch (LB) and a few sources   exhibit an extreme island state (van Straaten et al. 2003) at the lowest detectable luminosity. A vivid X-ray spectral transition is noted as atoll's moves across the various branches. The banana state is often found to be at high luminosity with a softer spectrum (Barret 2001; Church et al. 2014) whereas at lower luminosities, a hard tail is a key characteristic of the island state. These sources have two to five band-limited noise, peaked noise, 20-60 Hz quasi periodic oscillations (QPOs), typically show $\sim$ 200-1350 Hz QPOs and sometimes twin kilo Hertz (kHz) QPOs in their power density spectrum (PDS) (van Straaten et al. 2005; van der Klis 2006 and references therein).

On the other hand Z sources X-ray radiation close to Eddington luminosity (0.1-1 L$_{E}$) and trace a Z-like shape in the HID and CCD (Hasinger \& van der Klis 1989; van der Klis 2006). The Z shape is broadly divided into three branches where the energy spectrum and power spectrum systematically changes viz. horizontal branch (HB), normal branch (NB) and flaring branch (FB) with two vertexes i.e. at HB to NB and NB to FB transitions. Further classification depicts two subgroups among Z sources, Cyg X-2 like (Cyg X-2, GX 340+0 and GX 5-1) and Sco X-1 like (Sco X-1, GX 17+2, and GX 349+2) based on the shape and orientation of their branches (Kuulkers et al. 1994, 1997). XTE J1701-462 is an arch-type source exhibiting characteristics of both Z sources (Cyg X-2 and Sco X-1 like) as well as atoll sources (Homan et al. 2007; Lin et al. 2007). At high luminosity, XTE J1701-462 displayed Sco X-1 like variability followed by Cyg X-2 like variability at relatively lower luminosity and atoll like behavioral at decay phase. In general the variability across the Z shape is often considered to  increase in the mass accretion rate ($\dot{M}$) from HB to FB whereas an opposite scenario was proposed (Church et al. 2008; Jackson et al. 2009). On the contrary, it was suggested that $\dot{M}$ is constant along the branches and Z track could be due to instabilities either due to radiation pressure at inner radius/boundary layer or possibly caused by different solutions of accretion flows (Homan et al. 2002, 2010; Lin et al. 2007). High resolution spectral data unveil that inner disk radius appears to be close to the neutron star (NS) at different luminosities using {\it NuSTAR}, XMM-{\it Newton} and {\it Suzaku} (Cackett et al. 2012; Chiang et al. 2016; Ludlam et al. 2017). The PDS is often associated with a 10-60 Hz QPO at HB, 5-8 Hz QPO at NB (quite different from HB QPOs) along with kHz QPOs. It is found the QPO frequency increases from $\sim$ 10 Hz at the top of HB to 60 Hz at NB and a 5-8 Hz QPO is suddenly appears which is probably connected to a puffed-up disk (Titarchuck et al. 2001; Sriram et al. 2011), though a proper mechanism is still unknown.

Various models can explain the spectra of Z sources at different branches leading to a model degeneracy and because of this it is difficult to constrain the geometry of accretion disk. The soft (thermal) spectral component of Z source is modeled by a single temperature black body or a multicolor disk black body (Mitsuda et al. 1984) originating from the accretion disk while the non-thermal hard component is associated to a corona/Compton cloud, a jet (Di Salvo 2000, 2002; Agrawal \& Sreekumar 2003; Falanga et al. 2006; Lin et al. 2009) or bulk motion Comptonization (Titarchuk \& Zannias 1998; Farinelli et al. 2007). These models assume that hard X-ray emission is associated to a quasi spherical corona located in the inner region of the accretion around the NS surface. However in the accretion disk corona model wherein corona is extended over the Keplerian disk is often used to unfold the spectra of Z sources using a cut-off power-law model (Church \& Balucinska-Church 2004; Church et al. 2008, 2014). Radio emission associated with jets were observed at HB and NB (Migliari \& Fender 2006; Migliari et al. 2007; Fender et al. 2007) and in-fact a jet model can explain both spectral and timing variability of Z sources (Reig \& Kylafis 2015, 2016) where soft photons from disk/boundary layer are comptonized up-scattered in the jet.   

Cross correlation function (CCF) studies provide important evidences on the geometry and location of various soft and hard X-ray emitting regions in  accretion disk. Low Fourier time scale of 16/32 s anti-correlated hard delays of the order of 10 --1000 s were reported in a few BHBs (Choudhury et al. 2005; Sriram et al. 2007, 2009, 2010) indicating towards a truncated accretion disk geometry. Similar delays both anti-correlated hard and soft X-rays were reported for Z and atoll sources (Cyg X-2, Lei et al. 2008, GX 5-1, Sriram et al. 2012; 4U 1735--44, Lei et al. 2013; 4U 1608--52, Wang et al. 2014). These studies especially in Z sources show that most of delays were detected when the sources were in HB and NB where a corona/jet is always present (Fender et al. 2009; Migliari et al. 2007, 2011). Based on a few 100 s delays, energy spectrum and PDS, Sriram et al. (2012) suggested that a truncated accretion disk geometry is needed to explain such longer delays. In case of the black hole binary GX 339-4, spectral studies associated with an anti-correlated delay indicated towards the condensation of corona in the inner region of accretion disk (Sriram et al. 2010; Meyer-Hofmeister et al. 2012). These detected delays were interpreted as the viscous/readjustment time scales in the inner region of accretion disk. At high frequencies (0.1-100 Hz), a few milli-second time lag (both soft and hard) were detected in Z sources, indicating toward the Comptonization process in the corona/jet (Vaughan et al. 1999; Kotov et al. 2001; Qu et al. 2004; Arevalo \& Uttley 2006; Reig \& Kylafis 2016) where soft photons are up-scattered causing the hard lags (van der Klis 1987; Qu et al. 2001;  Li et al. 2013). Milli-seconds soft lags either can be explained by a shot model where soft photons lags to hard photons (Alpar \& Shaham 1985) or in a two layer Comptonization model (Nobili et al. 2000).  

GX 17+2 is a burster Z source, located at the distance of 13 kpc (Galloway et al. 2008) with a possible spin frequency of 584 Hz (Psaltis et al. 1999; Homan et al. 2002; Belloni et al. 2002) with no confirmed optical counterpart. Using the XMM-{\it Newton} data, an inclination $\le$ 45$^{o}$ was reported which is supported by the fact that no X-ray dips have been observed (Cackett et al. 2010). First a detailed spectral and temporal study for GX 17+2 was performed by Homan et al. (2002) and argued that $\dot{M}$ is not the unique parameter to explain the flux/luminosity evolution on the Z track and found to be constant in GX 17+2. A similar scenario was discussed based on the multi-component spectral model and illustrated the evolution of Z track with a constant $\dot{M}$ along with three different process viz. Comptonization in HB, existence of slim disk transition from a standard thin disk as the source approaches to NB and rapid decrease of thin disk radius on FB (Lin et al. 2012). A broad iron line and a low inclination of about 25$^{o}$--38$^{o}$ was reported using {\it NuSTAR} data (Ludlam et al. 2017). Penninx et al. (1988) observed radio emission in GX 17+2, found to be strong in HB and weak in FB along the Z track. Migliari et al. (2007) performed simultaneous X-ray and radio observation and found similar results and strongly suggested that NB oscillation is connected to the jet. Based on the {\it NuSTAR} data, the inner disk radius was constrained at 10.3--13.0 km.   In order to constrain the geometry of inner accretion disk in GX 17+2, we report the CCFs between soft and hard X-ray light curves along with temporal and spectral studies using the PCA \& HEXTE data of RXTE satellite and NuSTAR satellite data.

\section{Data Reduction and Analysis}

The public archival data of {\it Rossi}-XTE satellite (Swank 1999) was analyzed for GX 17+2. The satellite consists of three primary instruments the Proportional Counter Array (PCA; Jahoda et al. 2006), the High Energy X-ray Timing Explorer (HEXTE; Rothschild et al. 1995; Gruber et al. 1996) and the All Sky Monitor (ASM; Levine et al. 1996). Among five proportional counter units PCU 2 data have been used for analysis as it was the longest observational and best calibrated unit. Energy dependent light curves were generated using the standard 2 mode data observations spanning more than 2000 s (e.g. Sriram et al. 2012). HEASOFT v6.8 software was used to reduce and analyse the data. Background light curves were generated using PCABACKEST V3.8 and appropriate background models were invoked (bright back ground $\ge$ 40 count s$^{-1}$ and faint background $\le$ 40 count s$^{-1}$). Single bit, generic bit and event mode data were used to study the PDS. For spectral analysis, PCU 2 data were used and calibration uncertainties were ascertained by adding 0.5\% systematic errors. Spectra from HEXTE cluster A data were extracted in 15-50 keV. For generating source and background spectra, {\it hxtback} command was utilized, with a rocking interval of 32 s and were corrected for dead-time corrections using hxtdead (e.g. Sriram et al. 2013). Various models were used to unfold the spectra made available in XSPEC V12.7.1 (Arnaud 1996). HID was generated using the data collected in the energy domain of 6.2-8.7 keV, 8.7-19.7 keV and 2.0-19.7 keV with bin size of 32 s after carefully selecting the channels using RXTE gain epochs and gain change effects.

The Nuclear Spectroscopic Telescope Array Mission ({\it NuSTAR}) data analysis was performed for both Focal Plane Modules A and B i.e. FPMA and FPMB instruments (Harrison et al. 2013) by using a 120” circular region (similar to Ludlam et al. 2017) for background subtracted energy dependent light curves and for source and background spectrum. Background subtraction was done by using the region of same radius away from the source. The NuSTAR Data Analysis Software (NuSTARDAS) which is integrated in the HEASoft package was used for data extraction. The NuSTAR pipeline nupipeline was used to run all the data processing tasks in sequence. Nupipeline was used to perform data processing until the first 2 stages which is data calibration and data screening in order to produce cleaned event files. Upon producing the cleaned event files, the nuproducts task was used for products extraction (stage 3). DS9 was used for region selection and the nuproducts task was run by providing appropriate source and background region files, along with appropriate bin sizes and energy ranges for background subtracted light curve extraction  and the same nuproducts task was used to obtain source and background spectrum, with the corresponding response and ancilliary files. Upon obtaining both the FPMA and FPMB source and background spectrum, along with the ancilliary response matrices, they were combined using the ADDASCASPEC command. ADDRMF was used to combine and produce a single redistribution matrix file (RMF). The GRPPHA command was used to group the combined spectra to have a minimum of 25 counts per bin.
As data span over a wide range, the observation span of RXTE and NuSTAR satellite data has been indicated on the ASM lightcurve in Figure 1.

\section{Timing Analysis}

CCF studies were performed using soft (2-5 keV, 3-5 keV) and hard X-ray (16-30 keV) light curves obtained from the PCU2 of {\it RXTE} and {\it NuSTAR} data. It was found from the RXTE data that whenever the source was in FB, the CCF does not show any delay with a positive CCF. In HB and NB (Figure 2, top), we found a positive correlation with no delays, correlation with soft and hard delays, anti-correlation with zero delays, anti-correlation with soft and hard delays (Figure 3 \& Table 1). Hard/soft delay means that hard X-ray photons delayed to the soft/hard photons. It was found that delays were seen on a time scale of tens to a few hundred seconds and as such systemic trend is not observed from HB to NB. Similar behavior has been previously observed in Z type sources (Lei et al. 2008; Sriram et al. 2012; Ding et al. 2016). It is noted that whenever the anti-correlated delays occurred, CCF coefficients are often centered around $\sim$0.4--0.6 suggesting that hard X-ray photons are not linearly responding to soft photons. We found delays during which radio observation evidently point towards a jet (Migliari et al. 2007; see their Figure 1 \& 2). In ObsID 700023-01-01-01, the first section clearly shows a correlated hard delay of 100$\pm$28 s (radio flux was high), followed by an anti-correlation and ending with an anti-correlated soft delay of  --168$\pm$42 s (relatively low radio flux). If the NS surface/disk and hard photons are arriving from the base of the jet, then decrease in size of the base of jet would tend to decrease the hard flux too. The $\sim$100 s delay would be a time scale corresponding to change in the size of the base of jet.

{\it NuSTAR} data also revealed an anti-correlated hard X-ray delay of the order of $\sim$ 113$\pm$51 s with a CCF coefficient centered around $\sim$--0.6 at HB/NB branch (Figure 2, bottom) along with uncorrelated CCFs (Figure 4 \& 5). The soft and hard energy band X-ray light curves clearly show that soft and hard flux is oppositely varying. In a truncated accretion disk scenario, the delay can be explained if initially the disk truncated away and due to radial movement of disk, it cools the inner corona resulting in relatively low hard flux. Or if the accretion disk is present close to the NS surface at about 1.03--1.30 ISCO as reported by Ludlam et al. (2017) then the observed hard delay could be a gradual decrease in the size of the region responsible for the hard flux, which could be the corona varying in size.

\section{Power Density Spectrum}
Power density spectra were obtained for ObsID 20053-03-03-01 and 70023-01-01-01 for three sections of the light curves (left and right panels of Figure 6).  As discussed previously, ObsID 70023-01-01-01  shows relatively high degree of associated change in the X-ray and radio flux which is also associated with delay in CCF. It is observed that ObsID 20053-03-03-01 shows similar kind of light curve variation and hence we performed a comparative study. Both the PDSs are associated with HB with hardness ratios (HR) varying from 0.060-0.037 for 20053-03-03-01 and 0.046-0.064 for 70023-01-01-01.

 In ObsID 200053-03-03-01, QPOs were found to vary from $\sim$ 32 Hz (section A) to $\sim$ 24 Hz (section B \& C) and associated CCFs show soft X-ray delays of $\sim$- 207 s in section A, $\sim$ -668 s in section B and $\le$-100 s in section C (Figure 3b). Whereas in the other observation, 70023-01-01-01, a $\sim$26 Hz QPO along with a harmonic was present in all the sections within error bars. From the CCF analysis, we observed a hard delay of $\sim$ 114 s in section A and a soft delay of $\sim$ --49 s in section C (Figure 3k and Table 1). Although in both observations a $\sim$26 Hz QPO was present, delays were found to drastically change. Assuming that observed delays are readjustment time scales associated with truncated accretion disk geometry, these results indicate that a QPO alone cannot constrain the accretion disk geometry.

\section{Spectral analysis}   
To constrain and estimate spectral variation during delays, we investigated the ObsID 700023-01-01-01, for each section of the light curve (Figure 7) during which radio flux changed considerably along with X-ray flux. The source is in the upper horizontal branch during this observation (see Figure 2, Top). Section A, with a delay of 114 s, has a CCF of $\sim$ 0.6 and Section B shows a CCF of -0.3 with no delay, while section C shows a delay of -168 s having a CCF of -0.45 (Figure 3k). We unfolded spectra by two different models, first with a thermal Comptonization model viz. {\it wabs(BB+Gaussian+ nthcomp)} (Zdziarski et al. 1996), where we assumed that seed photons arriving from the NS and/or boundary layer are Comptonized in the Compton cloud. It was shown that such a thermal comptonization is needed for this source (Di Salvo et al. 2000), however we opted for {\it nthcomp} instead of {\it CompTT} model (Titarchuck 1994). We found that a Gaussian emission line is needed and its centroid energy was fixed at 6.5 keV and hydrogen equivalent column density N$_{H}$ was fixed at 3 $\times$ 10$^{22}$ cm$^{-2}$.  In nthcomp model, there are five parameters viz. the asymptotic power-law photon index, electron temperature (kT$_{e}$), seed photon temperature, {\it input type, kT$_{soft}$} which decide which seed photon will Compton up-scatter either it could be from disk or black body, and the normalization. The black body seed photons were chosen to fit the continuum and  kT$_{soft}$ parameter was allowed to vary. It was noted that this model gave a reasonable fit without any addition of a power-law component in the energy range of 3-50 keV, however a power-law component was needed in the analysis by Migliari et al. (2007). Unfolded spectra along with their model components are shown in Figure 7. The asymptotic power-law index, $\Gamma_{nthcomp}$ = 2.34$\pm$ 0.24 and the electron temperature kT$_{e}$ = 4.18 $\pm$ 0.85 keV in the first section were found to increase in the second and third sections ($\Gamma_{nthcomp}$ = 2.80$\pm$0.16, kT$_{e}$=7.59 $\pm$ 2.01 keV) (see Table 2). 
 For  ObsID 20053-03-03-01, we unfolded the spectra similarly  using the model {\it wabs(BB+Gaussian+nthcomp)}. A few of the parameters were fixed as discussed above. We found that kT$_{e}$ increase from 2.15$\pm$0.85 keV (section A) to 5.82$\pm$2.20 keV (section C) and $\Gamma_{nthcomp}$ varies from 2.28$\pm$0.08 to 2.67$\pm$0.22.

In the next attempt we used accretion disk corona configuration (Church et al. 2012) and fitted the spectrum with a model {\it wabs(BB+Gaussian+cutoffpl)}, where BB is black body emission from the NS and {\it cutoffpl} is the cut-off power-law associated with Comptonization in an extended corona over the disk. We found that all the parameters varied from the first to other sections. Church et al. (2012) fixed the power-law index $\Gamma$ of {\it cutoffpl} at 1.7, however we allowed it to vary. For the first section, $\Gamma$ = 1.13$\pm$0.24 changed to become a relatively steeper $\Gamma$ =1.76$\pm$0.17, $\Gamma$ =1.84$\pm$0.16 in the second and third sections and the cut-off energy E$_{c}$=5.31 $\pm$ 1.01 keV varied to higher values, E$_{c}$=8.32 $\pm$ 1.15 keV, E$_{c}$=8.93 $\pm$ 1.45 keV.  Same model was used for unfolding the spectra of ObsID 20053-03-03-01. $\Gamma$ is found to be 0.93$\pm$0.16 in the first section, which varied to a value of $\Gamma$ =1.43$\pm$0.11 in the second section to a slightly lower value of $\Gamma$ =1.30$\pm$0.12 in the third section. The cut-off energy E$_{c}$ went from 3.40 $\pm$ 0.55 keV in the first section to a relatively higher value of E$_{c}$=5.94 $\pm$ 0.65 keV in the second section and  E$_{c}$=5.64 $\pm$ 0.73 keV in the third section.

 Radio flux density is found to vary approximately from 4.5-1 mJy at 8 GHz and 4-2 mJy at 5 GHz (Migliari et al. 2007) during the observation duration of ObsID 70023-01-01-01 (see Figure 2 in Migliari et al. 2007). From both spectral fit attempts, it is clear that when the radio and X-ray fluxes were high (first section of the light curve) the kT$_{e}$ was low with a relatively hard power-law component and the state of gradual fall of both radio and X-ray fluxes is associated with a corona of relatively high temperature electron distribution kT$_{e}$. Such a configuration of corona i.e. a compact low temperature optically thick corona is the characteristic of a steep power-law state/intermediate state in black hole X-ray binaries (Done \& Kubota 2006; Done et al. 2007; Sriram et al. 2007, 2016) where often a transient jet is switched on/off (Fender et al. 2009).

 Another observation (ObsID 80022-01-05-00) was analyzed, where soft and hard X-ray light curves are anti-correlated for almost 23 ks and a steep variation has occurred at $\sim$18 ks (soft X-ray is decreasing but hard X-ray is increasing, see Figure 3q).  During this observation, the source was in the lower horizontal branch and upper normal branch. Same models as above were invoked to fit the 3-40 keV spectrum (as $>$ 40 keV S/N was low and hence was not considered) of each section (see Table 3 and Figure 7). The soft seed photon parameter kT$_{soft}$ was tied to kT$_{BB}$. It was found that the electron temperature is kT$_{e}$ $\sim$ 3 keV in all sections and no significant variations are observed in {\it BB} model parameters. A systematic residual is observed in the last section around $>$30 keV which can be modeled with a power-law component with an index, $\Gamma$=1.10 resulting in an improvement in the fit i.e.$\chi^2$/dof = 65/53 to $\chi^2$/dof = 36/53, F-test probability=4.4 $\times$ 10$^{-7}$ (see Table 3). Moreover it can be seen that the hard count rate increased in the final section of the light curve. This probably suggests a sudden appearance of jet in the inner region of accretion disk.

\subsection{Spectral comparison ObsID 70023-01-01-01 and 20053-03-03-01 }

A CCF and PDS study of both the observations indicated that different geometries are possible for the associated similar QPOs. We unfolded the PCA spectra for all the sections for both the observations using a model wabs(diskbb+Gaussian+nthcomp) (see Table 4)  with kT$_{soft}$ parameter of {\it nthcomp} model was tied to  kT$_{in}$ of {\it diskbb model} in order to constrain the fit. For the section A of 70023-01-01-01, the disk temperature was found to be kT$_{diskbb}$=1.47$\pm$0.09 along with the disk normalization N$_{diskbb}$ = 32 $\pm$7 whereas in section B and C kT$_{diskbb}$=1.09$\pm$0.16 was similar to each other along with N$_{diskbb}$ $\sim$ 129 and $\sim$145. The thermal comptonization parameters remained similar within uncertainties. Clearly with respect to the first section, the inner disk front has moved away from the neutron star in other sections. It is to be noted that Migliari et al. (2007) observed that radio flux was high in section A in comparison to the rest. In ObsID 20053-03-03-01, in section A, kT$_{diskbb}$=1.25$\pm$0.10 and N$_{diskbb}$$\sim$127, in section B \& C kT$_{diskbb}$=0.92$\pm$0.10 and N$_{diskbb}$$\sim$330 and the thermal comptonization parameters did not vary significantly. During both the observations thermal comptonization parameters were found to be similar i.e $\Gamma_{nthcomp}$ $\sim$ 1.8 and kT$_{e}$ $\sim$ 3.30 keV (Table 4).   

In one of the observations (ObsID 70023-01-01-01), we found a correlated hard X-ray delay of the order of 114 s with a relatively high correlation coefficient CC = 0.62$\pm$0.11. In order to ascertain the responsible spectral parameter causing the delay, we studied PCA spectra for the initial (part A) and final (part B) 300 s parts of the first section (see Figure 3k). We used the same model as discussed above. We find that with only simple thermal comptonization model having a Gaussian component provides a reasonable fit ($\chi^2$/dof=32/42, {\it BB+nthcomp}, $\chi^2$/dof=33/44, {\it nthcomp}). 
It was noted that $\Gamma$ has steepened with a 1$\sigma$ significance from 1.87$\pm$0.02 to 1.98 $\pm$0.02 whereas other parameters are consistent with no significant variation (Table 5). The steepening of $\Gamma$ indicates that the temperature distribution of electrons in corona has shifted towards a softer distribution.

 \subsection{NuSTAR spectrum of GX 17+2}
We studied the {\it NuSTAR} spectrum of the last section of the light curves where a delay was observed (Figure 5). We unfolded the spectrum invoking two different models wabs(BB+Gaussian+nthcomp) and wabs(diskbb\\
+Gaussian+nthcomp) (Figure 8). The kT$_{soft}$ parameter of {\it nthcomp} was tied to the soft seed photons arising from the neutron star surface i.e. kT$_{BB}$ of BB model or inner region of the disk, kT$_{in}$ of {\it diskbb} model. From both the models, it was found that the associated corona has $\Gamma$ $\sim $ 2.0 and electron temperature kT$_{e}$$\sim$3.0 keV (Table 6). It is be noted that   
inner disk radius R$_{in}$ = 14$\pm$1.0 which is close to the value reported by Ludlam et al. (2017). This suggests that the inner disk front is close to the inner stable circular orbit (ISCO).

\section{Discussion and Results}

In GX 17+2 the CCFs analysis clearly show the correlated and anti-correlated soft and hard X-ray delays of a few tens to hundred seconds which are a characteristic feature often observed in Z and atoll sources (Lei et al. 2008 and Sriram et al. 2012). It is noted that the observed delays in the source was associated to the HB and NB during which a radio jet was observed to be present in a few observations (Migliari et al. 2007). 

The spectral results suggest that the coronal properties have changed from a higher radio and X-ray flux to a low flux state. We found that disk normalization varied from section A to B \& C (see Table 4). In order to properly constrain the inner disk radius, {\it nthcomp} model parameters were frozen. Inner disk radii were derived using a relation  R$_{in}$ (km) = 1.2 $\sqrt{(N / cos i)}$ $\times$ D / 10 kpc (Reynold \& Miller 2013) and we noticed that R$_{in}$ has changed from 9.3 km (section A) to 18.7 km (section B)  (assuming {\it i} to be 25$^{\circ}$ and 35$^{\circ}$; Ludlam et al. 2017) but QPO frequency did not vary (see the right panel in Figure 5). Moreover, Migliari et al. (2007) showed that during this observation radio flux has varied consistently. In truncated accretion disk geometry, the QPO varies along with R$_{in}$ (e.g. Done et al. 2007) however in this observation it is not so. This probably suggests that truncated disk front can be situated at any location. A similar spectral analysis for ObsID 20053-03-03-01 for section A, B and C indicated that R$_{in}$ moved from 18 km (section A) to 31 km (section B \& C) whereas QPO varied from 32 Hz to 24 Hz (Table 4 and Figure 6), which can be explained from the truncated accretion disk scenario. Instead of diskbb model, if BB model is used, we observe correlated changes of electron temperature of nthcomp model with QPO frequency i.e. as QPO frequency decreases the corona temperature increases (see Table 4). We noticed that in ObsID 20053-03-03-01 when QPO is at 24 Hz,  R$_{in}$ is around $\sim$ 30 km whereas in ObsID 70023-01-01-01, R$_{in}$ varied from 9--20 km while QPO remained at $\sim$ 26-24 Hz.\footnote{Spectra were carefully corrected for gain variations during these epochs.}   

A few observations (Figure 3 \& 6) exhibit delays in their CCFs although the QPO frequency was not found to vary. For example, the observation 20053-03-03-01, in section A shows a QPO at 32 Hz while a QPO at $\sim$24 Hz is observed in both section B \& C (Figure 6). In respective CCFs, a soft X-ray delay varying from 207 s to 668 s from section A to B was noticed, while the delay is almost $<$100 s in section C. Assuming that the QPO is indicative parameter of the truncation radius, it can be concluded that the disk radius has changed from section A to B and the associated delays were found to be varying. Whereas in section B \& C, significant variation of delays were noticed although QPO remained the same. This probably indicates that in section B there is an appreciable change in the disk--corona structure and this structure readjustment has settled down, hence relatively smaller delay in section C.

\section{Constraining the Coronal Height} 

 It was assumed that the readjustment time scale of disk in the inner region is the causative factor of the observed delays of a few tens to hundred seconds between 2-5 keV and 16-30 keV in GX 17+2 , but this was found to be only a few tens of seconds. Even the readjustment time scales of the vertical structure of disk due to the radiation pressure would also result in delays of a few seconds only. Hence we could conclude that the observed delays could not be just due to the readjustment time scales of the inner region of the disk but corona is also changing its size and structure. Thus we must consider the readjustment time scales of the coronal structure as well.

Assuming the observed delay is a combination of the readjustment time scale of the disk as well as corona, we can write,

\begin{equation}
t_{delay}=\frac{1}{v_{disk}}\Bigg[R_{disk}+ \frac{H_{corona}}{\beta}\Bigg]
\end{equation}

where the first term is the readjustment timescale in the disk (assuming a Shakura and Sunyaev disk) and the second term is the readjustment timescale in corona.

We assume that the readjustment velocity in the coronal region is $\beta$ times that in the disk where $\beta$ is $\le$ 1 ({\it v$_{corona}$=$\beta$v$_{disk}$}), since the coronal viscosity is less than the disk viscosity. We can rearrange the above equation to obtain the height of the coronal region,

\begin{equation}
H_{corona}=\Bigg[\frac{t_{delay} \dot{m}}{2 \pi R_{disk} H_{disk} \rho}-R_{disk}\Bigg] \times \beta \; cm
\end{equation}
where 
H$_{disk}$ = 10$^{8}$ $\alpha^{-1/10}$ $\dot{m}_{16}^{3/20} R_{10}^{9/8} f^{3/20} $ cm, 
$\rho$ = 7 $\times$ 10$^{-8}$ $\alpha^{-7/10}$ $\dot{m}^{11/20}$ $R^{-15/8}$ $f^{11/20}$  g cm$^{-3}$, f = (1-(R/R$_s$)$^{1/2}$)$^{1/4}$ 

Based on the above equations, we considered a delay of 207 s in the ObsID 20053-03-03-01 and considered R$_{disk}$ to be 21 km in keeping with the observed QPO frequency of 32 Hz using the relation $\frac{R_{in}}{R_g} \le 27\nu^{-0.35} \Bigg[{\frac{M}{2 M_{\odot}}}\Bigg]^{-2/3}$ (Di Matteo \& Psaltis 1999) and thus determined the coronal height to be $\sim$ 17 km ($\beta$=0.05) and 34 km ($\beta$=0.1) at that instant ($\dot{m}$ obtained from spectral fits; Table 4). Upon considering an observed soft delay of 668 s, based on the detected QPO frequency 24 Hz, we determine the coronal height to be $\sim$ 56 km ($\beta$=0.05) and 111 km ($\beta$=0.1). An $\alpha$ value of 0.1 was assumed for all the above calculations. We found that the inner disk radii R$_{in}$ values are similar to that obtained from QPO frequency inner disk radii (see equation 1 and  Table 4).  We also evaluated the H$_{corona}$ assuming the disk is at the last stable orbit i.e. R$_{disk}$ = 12 km, which resulted in higher values.

In the truncated accretion disk scenario, the coronal quasi-spherical structure is assumed to be present inside the truncation radius, which suggests that the coronal radius should be of the order of truncated inner disk radius. Substituting R$_{disk}$ = H$_{corona}$ in equation 1 (R$_{disk}$ being the inner truncation disk radius), one can constrain the $\beta$ factor which is of the order of 0.062--0.021 for a delay of 207 s \& 668 s (ObsId 20053-03-03-01 section A \& B) and $\beta$=0.125 \& 0.085 for a delay of 114 s and 168 s in ObsID 70023-01-01-01 (section A and C). As shown above that coronal readjustment time scale is primarily responsible for the observed delays, we argue that $\beta$ is playing a key role. Based on NuSTAR spectrum study, we found that the disk is close to ISCO $\sim$ 14 km. Then proportionally the coronal height shall change based on the above discussed scenario and we found that $\beta$=0.08 for a delay of 100 s.  Based on the above and this section we infer that the corona is varying independent of the location of the disk with different coronal readjustment time scales.

\section{Constraining the Disk Radius}
 The disk truncation radius has been constrained using multiple models, which include the Relativistic Precession Model (RPM; Stella et al. 1999) , the Transition Layer Model (TLM; Titarchuk \& Osherovich 1999) and the optically thin flow solution of the Advection Dominated Accretion Flow model (ADAF; Narayan \& Yi 1995a, b). Apart from those, radius was constrained by applying the relations for the different inner structures in the truncated disk scenario, which include the quasi spherical region of the radiation pressure dominated disk, the magnetosphere radius and also the boundary layer radius (Popham \& Sunyaev 2001). This section deals with each of the estimations separately.

\subsection{Disk radius from the RPM}
Based on the QPO analysis of the ObsID 70023-01-01-01, we found that QPOs are present with a centroid frequency $\sim$ 26 Hz along with a harmonic $\sim$ 51 Hz in all the sections. In general for NS LMXBs, the horizontal branch oscillations (HBOs) and kHz QPOs are related to the Keplerian orbital motion, precession of periastron and nodal precession of the perturbed orbit in the relativistic precession model (Stella \& Vietri 1999; Stella et al. 1999) and suggested that  R$_{in}$ $\propto$ $\nu^{-1/3}$, where R$_{in}$ is the inner disk radius and $\nu$ is QPO frequency (HBO). By studying the correlation among the LMXBs, Di Matteo \& Psaltis (1999) derived an upper bound on the inner disk radius

\begin{equation}
\frac{R_{in}}{R_g} \le 27\nu^{-0.35} \Bigg[{\frac{M}{2 M_{\odot}}}\Bigg]^{-2/3}
\end{equation} 

setting $\nu$=26 Hz, we found R$_{in}$ to be $\sim$ 23 km.

\subsection{Disk radius from the TLM}
We estimated inner disk radius using the transition layer model (Titarchuk \& Osherovich 1999; Osherovich \& Titarchuk 1999a, b). The study of Wu (2001) indicated that this model can better constrain the correlation between HBO and kHz QPO in Z sources. Using the characteristic relation between 
$\nu_{HBO}$, $\nu^{L}_{kHz}$ (lower kHz QPO) and  $\nu^{H}_{kHz}$ (upper kHz QPO)
\begin{equation}
\nu^{H}_{kHz} = \sqrt{(\nu^{L}_{kHz})^{2}+ \Bigg[\frac{\Omega}{\pi}\Bigg]^{2}}
\end{equation}

\begin{equation}
\nu_{HBO} = \frac{\Omega}{\pi} \frac{\nu^{L}_{kHz}}{\nu^{H}_{kHz}} sin\delta 
\end{equation}

where $\delta$ is the angle between rotational angular velocity ($\Omega$) and normal to the Keplerian oscillation ($\delta$=6.30 for GX 17+2; Wu 2001). Using the above equations and substituting into {R$_{in}$}/{R$_{g}$} $\le$ 220 $\nu^{L}_{kHz}$$^{-2/3}$ (M / 10 M$_{\odot}$)$^{-2/3}$ (Di Matteo \& Psaltis 1999), we get

\begin{equation}
\frac{R_{in}} {R_{g}} = 220\Bigg[\bigg(\frac{\Omega}{\pi}\bigg)^{-2/3} \bigg((\frac{\Omega\ \ sin \delta}{\nu_{HBO}\ \ \pi})^{2} - 1\bigg)^{1/3}\Bigg] \Bigg[{\frac{M}{10 M_{\odot}}}\Bigg]^{-2/3}
\end{equation}

We found R$_{in}$ to be  $\sim$ 21 R$_{g}$ ($\sim$ 44 km) for 26 Hz HBO.

\subsection{Disk radius from the ADAF solution}
The Compton cooling process plays an important role in structure of the corona where the ratio of ion temperature and electron temperature is important (Narayan \& Yi 1995a, b). It is found that for a small value of advection energy fraction f, the corona density increases towards the disk departing from the spherical structure (Meyer-Hofmeister et al. 2012). For optically thin flow of an ADAF solution (advection dominated accretion flow; Narayan \& Yi 1995b), r is given by

\begin{equation}
r=\Bigg[\frac{\tau}{8.27 {\alpha^{-1}} c_{3}^{-1} \dot{m}_{c}}\Bigg]^{-2} ,
\end{equation}

where $\tau$ is the optical depth of corona, $\alpha$ is the viscosity parameter, $\dot{m}_{c}$ the ADAF accretion rate in units of Eddington rate (1.39 $\times$ 10$^{18}$ g/s) and c$_{3}$ = 2/(5+2$\eta$)(the ratio of square of the sound speed to the Keplerian velocity), $\eta$=(1/f)$\times$((1.667-$\gamma$)/($\gamma$-1)), $\gamma$= (8-3$\beta$)/(6-3$\beta$), $\beta$, ratio of the gas pressure to the total pressure and r is the radius in units of Schwarzchild radius. For $\tau$ $\ge$ 1, f becomes low. Assuming f=0.2 , $\alpha$=0.5 and $\tau$=5 (from the relation given in Zdziarski et al. (1996) and using the $\Gamma$ in Table 3), $\dot{m}_{c}$=0.3 resulted in r $\sim$ 50 km. It should be noted that $\alpha$=0.5 was chosen (cannot be ruled out; Liu \& Taam 2013) and a lower value results in increase the value of radius ($\sim$ 317 km for $\alpha$=0.2). However c$_{3}$ can be constrained from the fact that H/R $\sim$ (2.5c$_{3}$)$^{1/2}$ (Narayan \& Yi 1995b) and assuming H/R = 2 (due to the radiation pressure; Ding et al. 2011), near 50\% of Eddington luminosity, we get r $\sim$ 28 km.

\subsection{Disk radius from constraining the inner quasi spherical region}
The radius of quasi spherical region of radiation pressure disk is given by R$_{sp}$ $\sim$ 32 $\dot{M}$$_{Edd}$ km (Ding et al. 2011). It is clear from these relations that for accretion rate $\dot{M}$ in the disk, there is a steep increase in the radiative pressure supported disk radius than the magnetically supported disk radius. In this case, we find that R$_{sp}$ $\sim$ 36 km assuming the total flux and $\sim$ 27 km for {\it nthcomp} flux for the section A. For section B, R$_{sp}$ $\sim$ 33 km and 22 km, section C $\sim$ 33 km and 22 km (Table 2). 

\subsection{Disk radius from constraining the boundary layer region}
A boundary layer (BL) is considered to be the region where the gas transitions from the disk to the star and it is formed to account for the change in angular momentum between the innermost portions of the accretion disk and the neutron star surface (Shakura \& Sunyaev 1988; Popham \& Sunyaev 2001). The luminosity of the BL directly depends on the velocity difference between the accreting material and the NS surface. In a NS, both the outer Centrifugal Barrier Supported Boundary Layer (CENBOL) and the inner normal BL take part in the accretion flow and emission processes. If the BL is responsible for the hard X-ray flux then the decrease in size of the layer could be the cause of the detected hard X-ray lag seen in the CCF.

It was shown that BL radius increases with mass accretion rate whereas height of the BL increase as H $\propto$ $\dot{m}$$^{0.45}$. We used the following equation to calculate the radius of BL (Popham \& Sunyaev 2001).

\begin{equation}
log(R_{BL} - R_{NS}) \sim 5.02 + 0.245 \Bigg[log\Big({\frac{\dot{M}} {10^{-9.85}\ \ M_{\odot} yr^{-1}}\Big)}\Bigg]^{2}
\end{equation}

In order to obtain the mass accretion rate, we used the equation L= $\frac{GM\dot{M}}{R}$ using M = 1.4 M$_{\odot}$ and R = 10 km.
 We found R$_{BL}$ $\sim$ 17.02 R$_{g}$  ( $\sim$35 km) for A, R$_{BL}$ $\sim$ 16.12 R$_{g}$ ($\sim$ 33 km) for B and R$_{BL}$ $\sim$ 15.85 R$_{g}$  ($\sim$ 33 km) assuming the total fluxes (ObsID 70023-01-01-01; Table 2). If we use thermal comptonization flux then R$_{BL}$ $\sim$ 13.73 R$_{g}$ ($\sim$ 29 km), $\sim$ 12 R$_{g}$  ($\sim$25 km) and $\sim$ 11.82 R$_{g}$ ($\sim$ 24 km). The BL radius decreased from section A to C along with decrease in the hard flux and this suggests that the disk is truncated at relatively larger radius during section A in comparison to section B's truncated radius. In a truncated disk geometry, as the inner disk front moves close to NS, the flux decreases and vice-versa. However this scenario is different if we consider the inner disk radius, since R$_{in}$ has a low value in section A (see Table 4). 

\subsection{Disk radius from constraining the magnetosphere}
The magnetic field of NS can also truncate the accretion disk and follows a relation viz. R$_{B}$ $\propto$ $\dot{M}$$^{2/7}$. The magnetospheric radius can be derived from the following equation given by Lamb (1973),

\begin{equation} 
r_{m} \sim 2.29 \times 10^{6} \mu^{4/7} \dot{m}^{-2/7} M^{-1/7} \;cm 
\end{equation}

Here $\mu$=15 is the magnetic dipole moment in units of 10$^{26}$ G cm$^{3}$, $\dot{M}$$\sim$100 (units of 10$^{16}$ g s$^{-1}$) and M is in units of 1.4 M$_{\odot}$. This led to an estimate of r$_{m}$ $\sim$ 28.8 km.

Three dimensional MHD simulation of magnetospheric accretion performed by Kulkarni \& Romanov (2013) led to a new dependence of the magnetospheric radius on the mass accretion rate as given by,

\begin{equation} 
r_{m} \sim 2.50 \times 10^{6} \mu^{2/5} \dot{m}^{-1/5} M^{-1/10} R^{3/10} \;cm 
\end{equation}

Here R is the NS radius in units of 10$^{6}$ cm.
Using this equation the radius was found to be  $\sim$ 29 km.

As given by the equation,

\begin{equation} 
\tiny
\Bigg[{\frac{R_{in}}{10 km}}\Bigg]=\Bigg[\frac{\mu }{0.56 k_{A}^{-7/4}\Bigg[{\frac{M}{1.4 M_{\odot}}}\Bigg]^{1.4} \Bigg[{\frac{f_{ang}}{\eta}}{\frac{F}{10^{-9}\ \ erg cm^{-2} s^{-1}}}\Bigg]^{1/2} {\frac{D}{3.5 kpc}}}\Bigg]^{4/7}
\end{equation}
where $\mu$ is the magnetic dipole moment in units of 10$^{25}$ G cm$^{3}$, k$_{A}$ the conversion factor from spherical to disk accretion, f$_{ang}$ the anisotropy correction factor, F is the flux, $\eta$ the accretion efficiency, D is the distance to the source in units of kpc and assuming B =1.5 $\times$ 10$^{9}$ G, k$_{A}$=1, f$_{ang}$=1 (for more discussion see Ibragimov \& Poutanen 2009), $\eta$ = 0.2,  R$_{B}$ (truncation radius due to magnetic field) was found to be  $\sim$ 31 km  for section A, B and C assuming the total flux (Table 2). However a change in the values of B varies the R$_{B}$ (eg. B = 0.5 results in R$_{B}$ $\sim$ 16--18 km).

The independent approaches to obtain the truncation radius resulted in a similar value suggesting that the primary source for X-ray flux is in the region of 12--17 R$_{g}$ during the HB of GX 17+2. The observed delays of the order of a few hundred seconds is the adjusting time scale of this region could be a corona possibly supported by both radiative pressure and magnetic field of the NS. Ludlam et al. (2017) reported that the inner disk is present close to the NS at about 1.03--1.30 ISCO (innermost stable circular orbit) $\sim$ 12-15 km and suggested that the boundary layer would be about of a size of 2 km. Coughenour et al. (2018) found that for Z source GX 349+2, the inner disk was found around $\sim$33 km similar to the value for Cyg X-2 (Mondal et al. 2018).

\section{Conclusion}
The energy dependent cross correlation functions (CCFs) of GX 17+2 associated with soft and hard X-ray delays clearly show delay of the order of a few hundred seconds during HB \& NB and are found to be highly correlated on FB with no delay. Highly correlated CCFs can be explained if both soft and hard X-rays are emitted from a confined zone, most probably arriving from the inner region of the accretion disk. The correlation coefficients of the anti-correlated and correlated CCFs of HB and NB are found to be lower than CCFs of FB indicating that the soft and hard X-ray emission are arriving from a broader region around the NS surface and hence the weaker correlation. It was observed that the CCF is highly positive with CC $>$ 0.8 with no delays in FB whereas this highly correlated feature of CCF is not seen in HB or NB where a lower CC has been observed with delays (e.g. Figure 3f). This is possibly due to the presence of an optically thick low electron temperature Compton cloud or a jet often associated with HB/NB with a characteristic of high variability. This scenario is closely matching with the results by Lin et al. (2012) where it was concluded that at constant mass accretion rate across the Z track,  geometry of Comptonization region plays a pivotal role in forming the HB  and a slim disk configuration up to the NB in the HID. Such a configuration of corona is often seen in black hole X-ray binaries (BHBs) where it is needed for launching a jet or an occurrence of jet line in HID (Done et al. 2007; Fender et al. 2009). During the episodic events of jet/ejection, it is often found that rapid transition occurs in BHBs where QPO types have rapidly varied or disappeared (e.g. Nespoli et al. 2003; Miller-Jone et al. 2012; Sriram et al. 2012, 2016). Z sources also display a rapid variability in QPO frequency during HB--NB or FB--NB transition (e.g. Casella et al. 2006). 

From the CCF, PDS and spectral studies, we arrive at the following conclusions.\\

1. For the first time, the CCF between 3-5 keV and 16-30 keV showed a hard X-ray delay of 113 s using {\it NuSTAR} observation of GX 17+2 in one of the sections of the light curve where spectral studies indicate that the disk is close to the last stable orbit (see Table 6) and similar value was reported by Ludlam et al. (2017) . But in other sections of the light curve, CCF was found to be uncorrelated.\\

2. Based on the analysis of RXTE observation of GX 17+2, we found soft and hard X-ray delays of the order of a few hundred seconds which are indeed a  characteristic feature associated with the HB and NB with a cross correlation coefficient arbitrarily spread in the domain $\sim$ 0.35 to 0.6. In the FB, the CCFs are strongly correlated with the coefficients $>$0.8. Though correlated and anti-correlated soft and hard delays were present across the HB and NB but such noticeable pattern of increase/decrease of delay values (i.e. higher/lower delays at the top of HB and lower/higher delays at the bottom of HB or NB) were not observed as the source traverses from the HB to NB or vice-versa.\\

3. In the case study performed for ObsID 70023-01-01-01 and 20053-03-03-01, we found that at similar QPO frequency $\sim$26 Hz the disk radii are different which indicates that disk truncation is not always correlated with QPO frequency. This result is evident from the section A and section C of ObsID 70023-01-01-01 where QPO has almost remained the same but disk radius varied from $\sim$10 km to $\sim$20 km.\\

4. From the observed delays in CCF, PDS and spectral studies, we conclude that corona is varying incoherently with the disk component, leading to different delays across the HB and NB. This is evident from ObsID 70023-01-01-01 and 20053-03-03-01 where for similar QPOs the observed delays were found to be different.\\

5. Coronal heights were constrained assuming that the delays are readjustment time scales of the inner truncated accretion disk region. Based on the delays, we found that on an average coronal heights are of the order of 20-35 km depending on the value of $\beta$ where $\beta$ is defined as the ratio of readjustment velocity of disk to the corona. Assuming that the coronal height is equal to the truncated radius we found that coronal readjustment velocities should be about $\beta$=0.06-0.12 times the readjustment velocities of disk. However it should be noted that the disk is close to the last stable orbit based on {\it NuSTAR's} spectrum. Assuming that the disk is always at the last stable orbit the detection of delay in {\it NuSTAR} light curve suggests that coronal properties viz. size or geometry varies independent of the disk location. But {\it NuSTAR} has yet to observe the spectral transitions in GX 17+2 similar to GX 5-1 (Homan et al. 2018) where a clear change in the disk radius is observed.  \\

6. We determined the disk truncation radius from various other methods as discussed above and found that it is of the order of 20-35 km.\\

It is evident from the above studies that both simultaneous timing and spectral observations are required in order to constrain the truncation radius of the accretion disk as well as to understand the corona geometry in Z sources and other X-ray binaries. Future long term observations from {\it Astrosat} and {\it NuSTAR} will be useful to constrain the timing and spectral variability and time-lag correlation at various spectral states in both Z and Atoll type neutron star sources during such delays, which could also constrain the size of corona and location of inner disk radius. \\


\begin{acknowledgements}

We thank the referee for the valuable and constructive feedback which led to the improvement of this paper. This research has made use of data obtained through the HEASARC Online Service, provided by the NASA/GSFC, in support of NASA High Energy Astrophysics Programs. This research has made use of the NuSTAR Data Analysis Software (NuSTARDAS) jointly developed by the ASI Science Data Center (ASDC, Italy) and the California Institute of Technology (Caltech, USA). K.S. acknowledges the financial support from DST-PURSE-II/1/17, Government of India. S. M. acknowledges the support from DST INSPIRE fellowship scheme (No: DST/INSPIRE Fellowship/[IF160388]). Part of the work has been carried out at Korea Astronomy and Space Science Institute.
\end{acknowledgements}
\begin{figure*}
\includegraphics[height=15cm,width=15cm, angle=270]{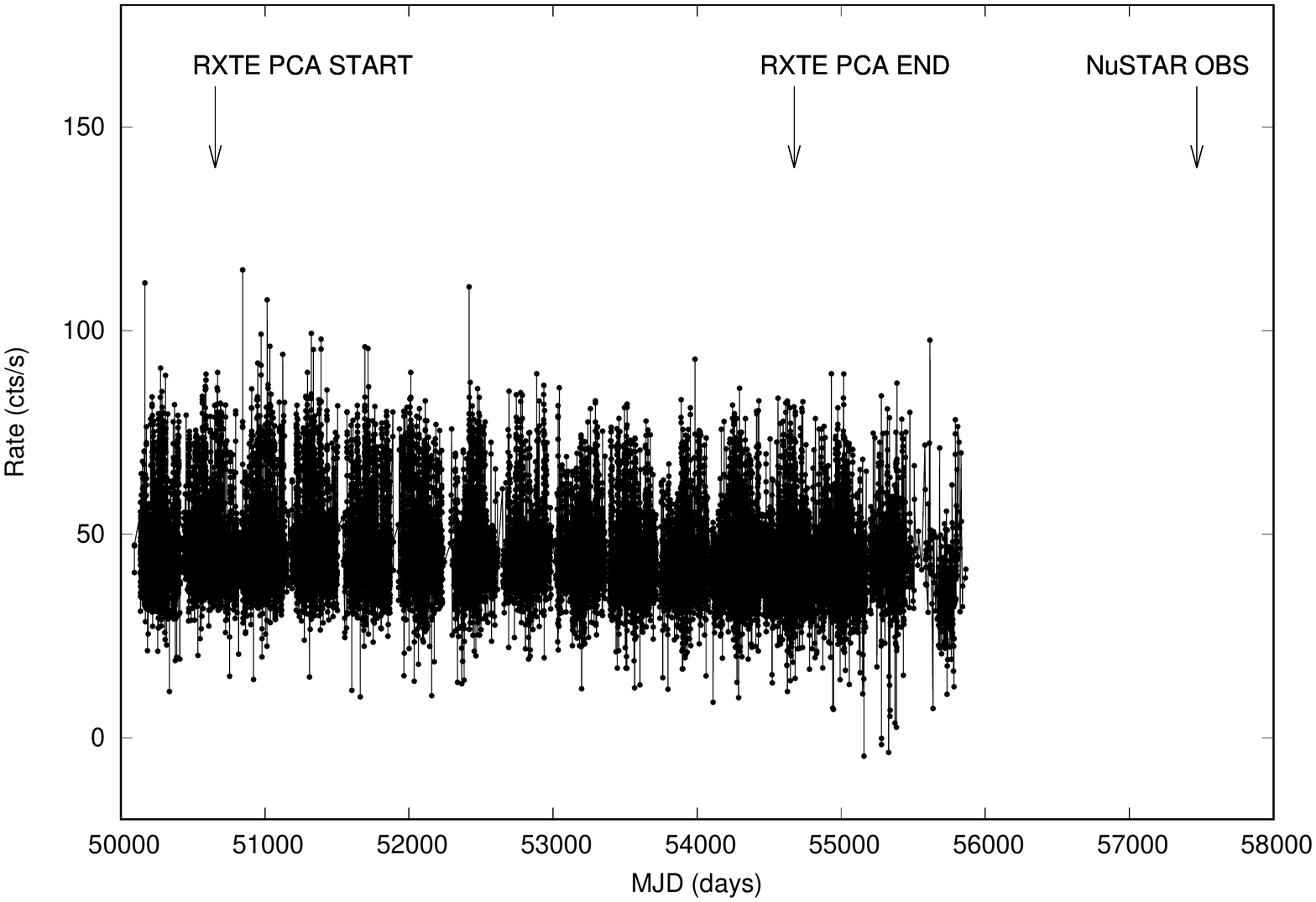}\\

\caption{RXTE ASM lightcurve of GX 17+2 with indicators for the start and end of RXTE PCA and NuSTAR observations.}
\end{figure*}

\begin{figure*}
\includegraphics[height=15cm,width=8cm, angle=270]{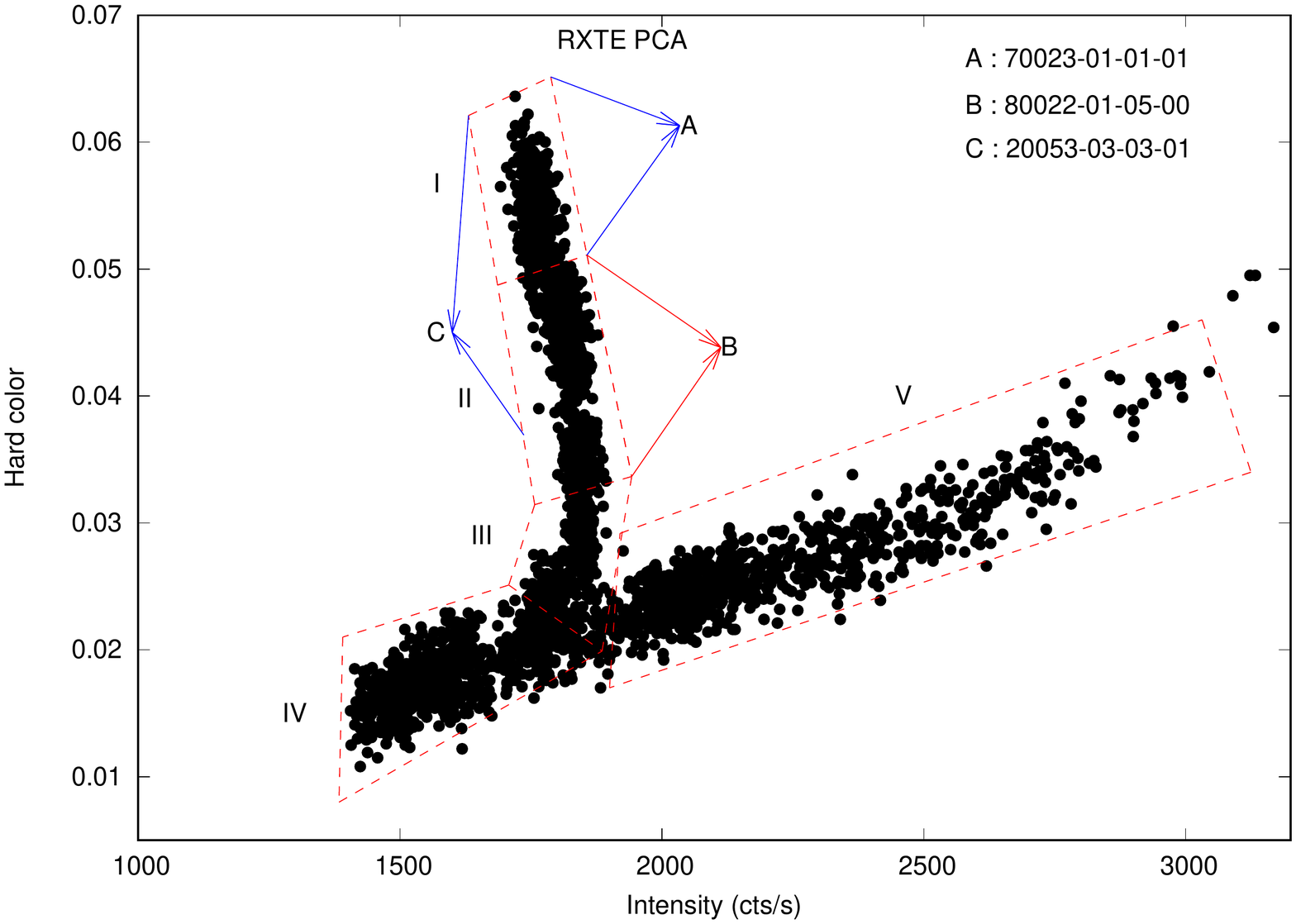}\\
\includegraphics[height=15cm,width=8cm, angle=270]{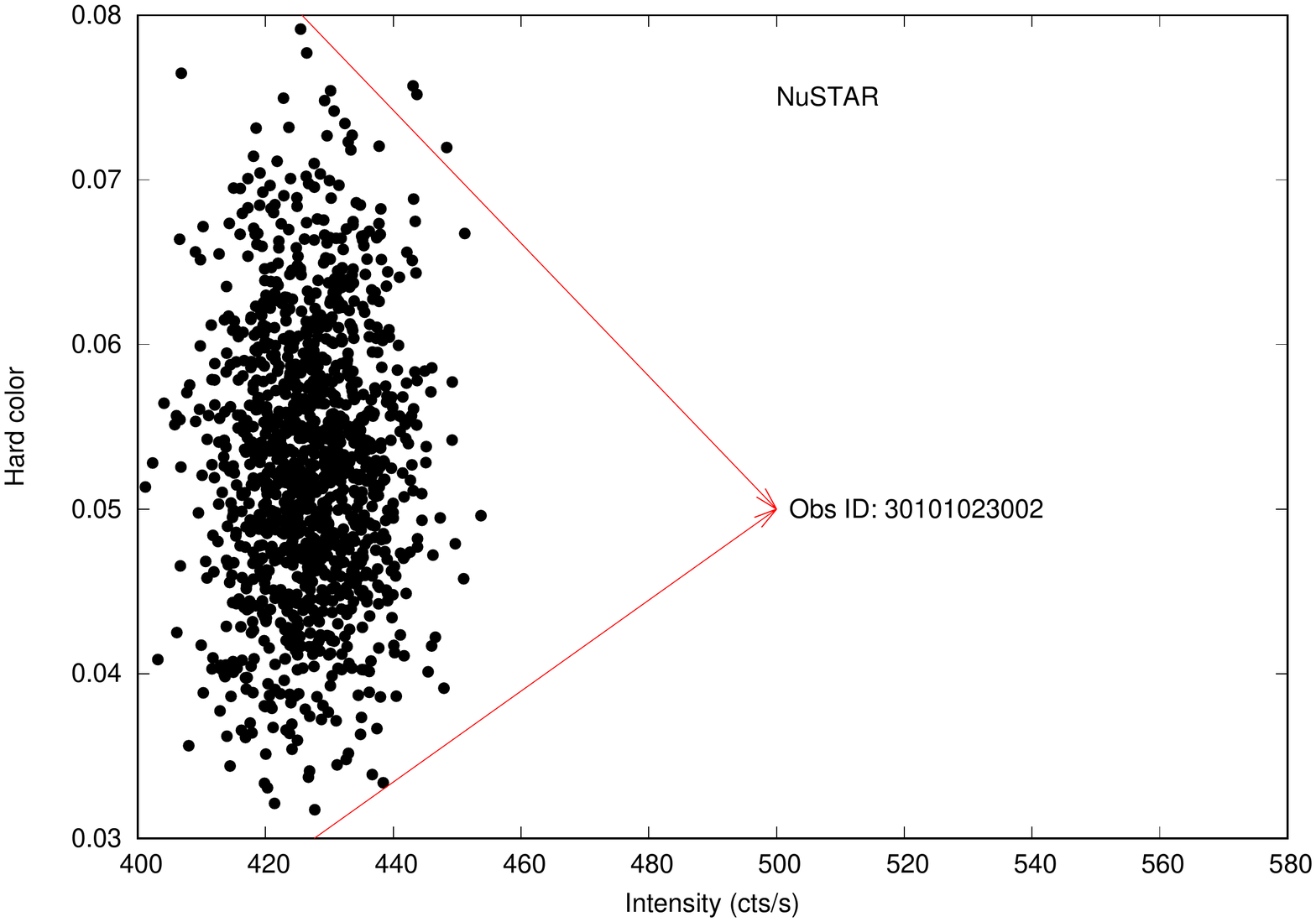}\\
\caption{Top: Z track for GX 17+2 using RXTE PCA observations. Hard color is defined as 16-30 / 2-5 keV and  Intensity is that in the 2-30 keV  range. Label A indicates the stretch of ObsID 70023-01-01-01, B indicates ObsID 80022-01-05-00 and C indicates ObsID 20053-03-03-01. Box marked I indicates upper HB, II is the lower HB and upper NB, III gives the lower NB, IV and V indicates the FB. 
Bottom: Same figure for NuSTAR observations. These points belong to the HB/NB region when compared to the HID of RXTE PCA data.}
\end{figure*}

\begin{subfigures}
\begin{figure}

\includegraphics[width=0.25\paperwidth,angle=270]{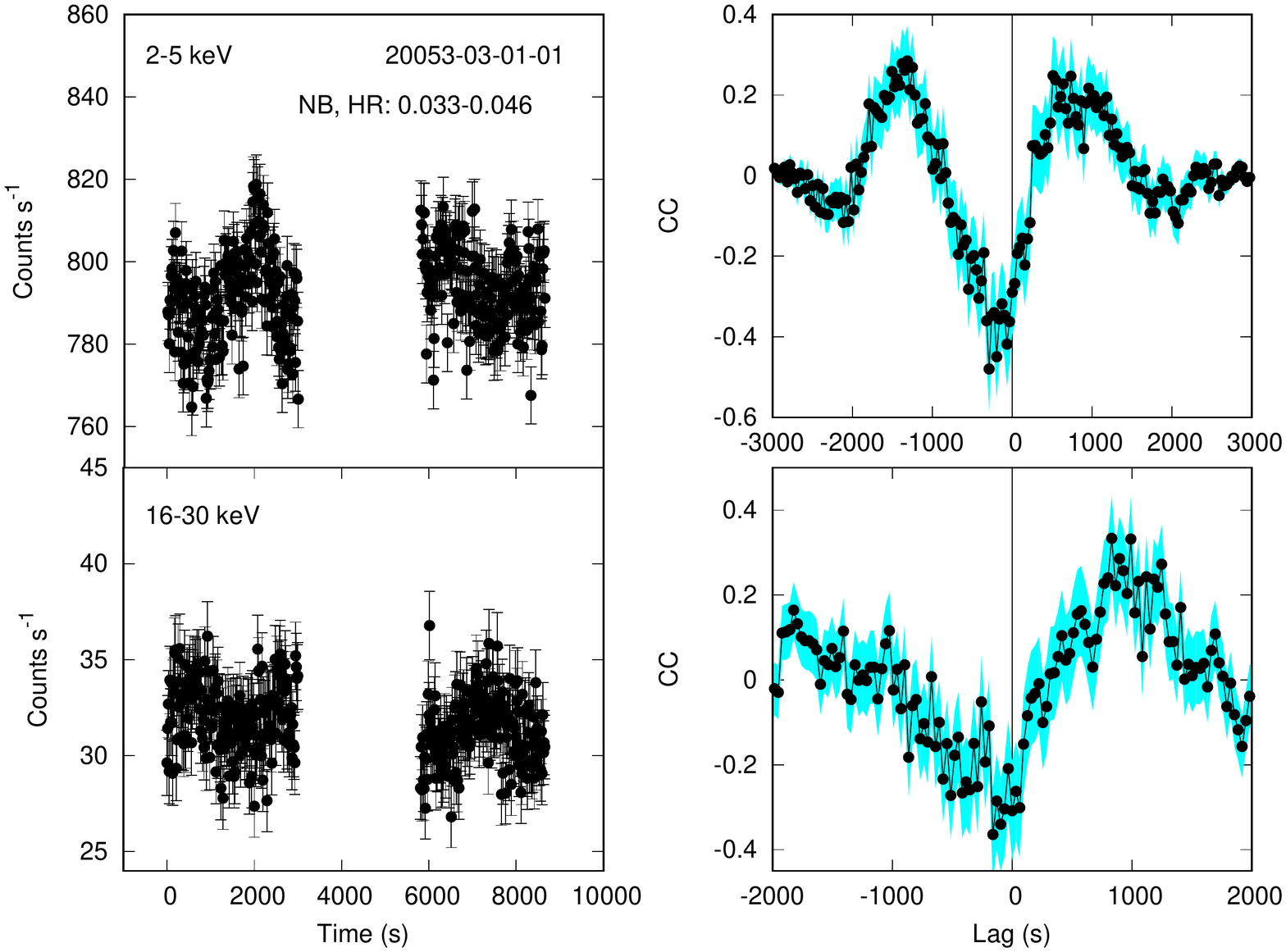} 
\caption{The background subtracted 32 s bin RXTE soft (2--5 keV) and hard X-ray (16--30 keV) light curves (left panels) for which correlated and anti-correlated X-ray lags are observed (right panels). The ObsID and energy bands are mentioned in the light curves. Right panels show the cross correlation function (CCF) of each section of the light curve and shaded regions show the standard deviation of the CCFs. }

\end{figure}

\begin{figure}

\includegraphics[width=0.25\paperwidth,angle=270]{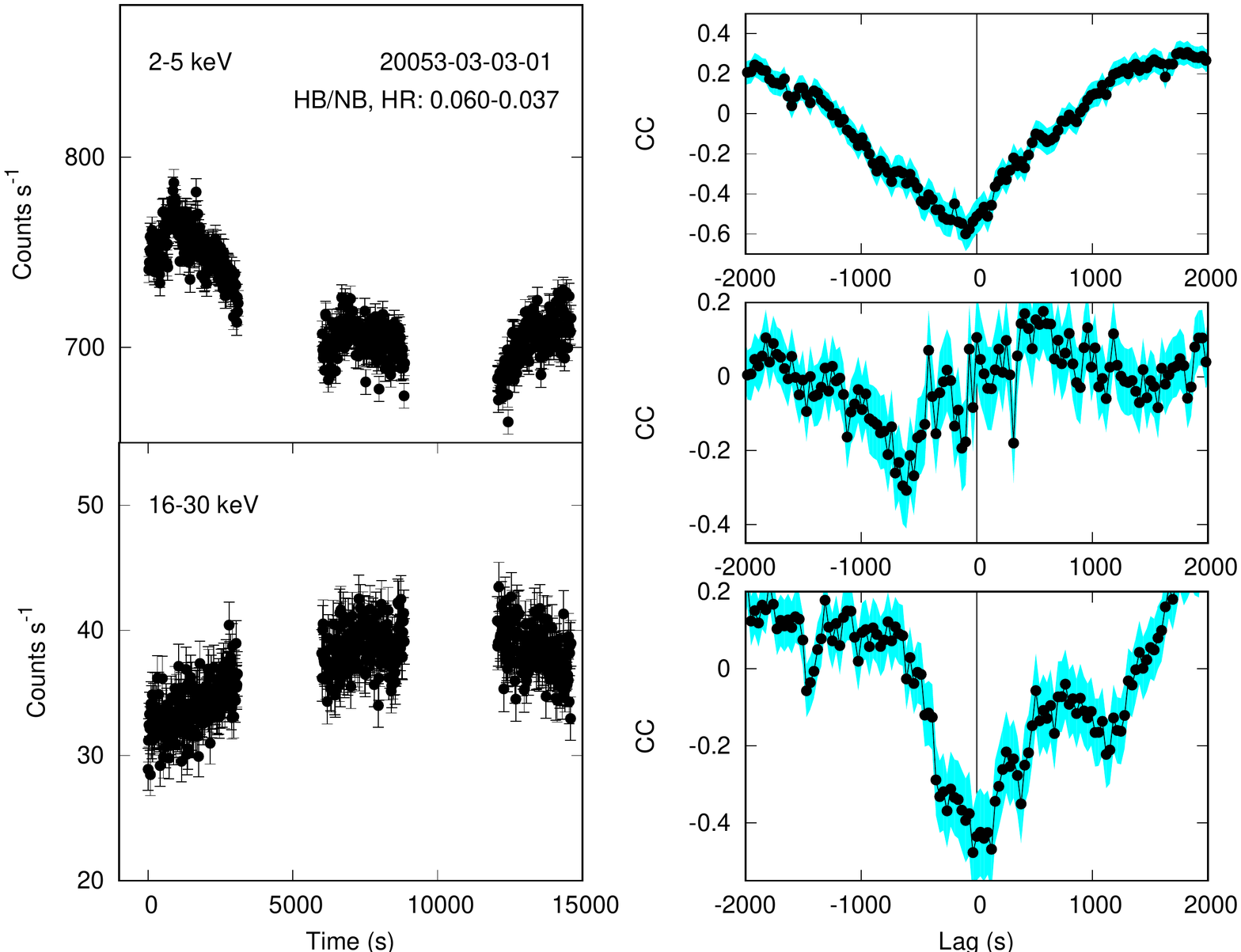}
\caption{Same as Figure 3a.}

\end{figure}

\begin{figure}
\includegraphics[width=0.25\paperwidth,angle=270]{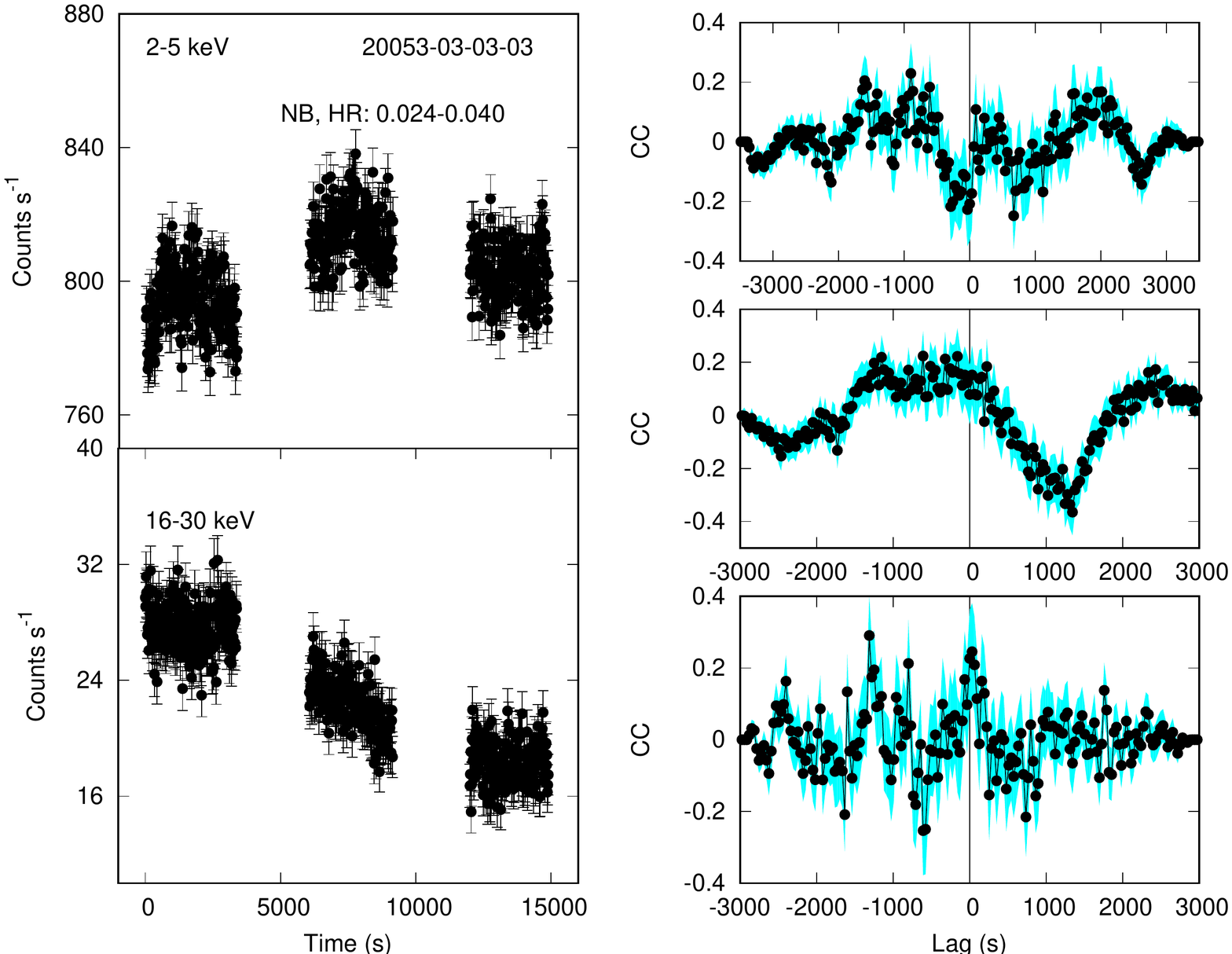}
\caption{Same as Figure 3a.}
\end{figure}

\begin{figure}
\includegraphics[width=0.25\paperwidth,angle=270]{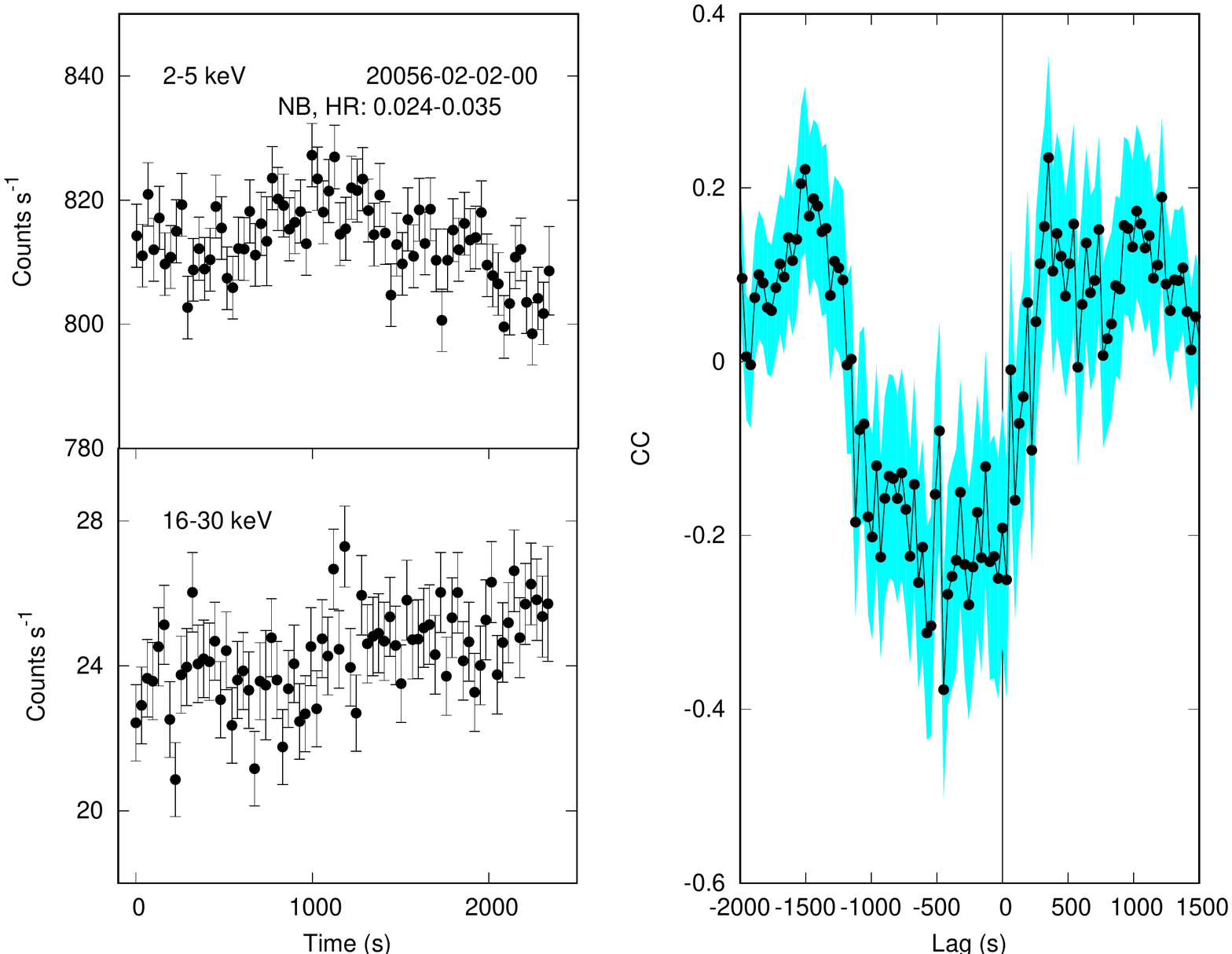}\\
\caption{Same as Figure 3a.}
\end{figure}

\begin{figure}
\includegraphics[width=0.25\paperwidth,angle=270]{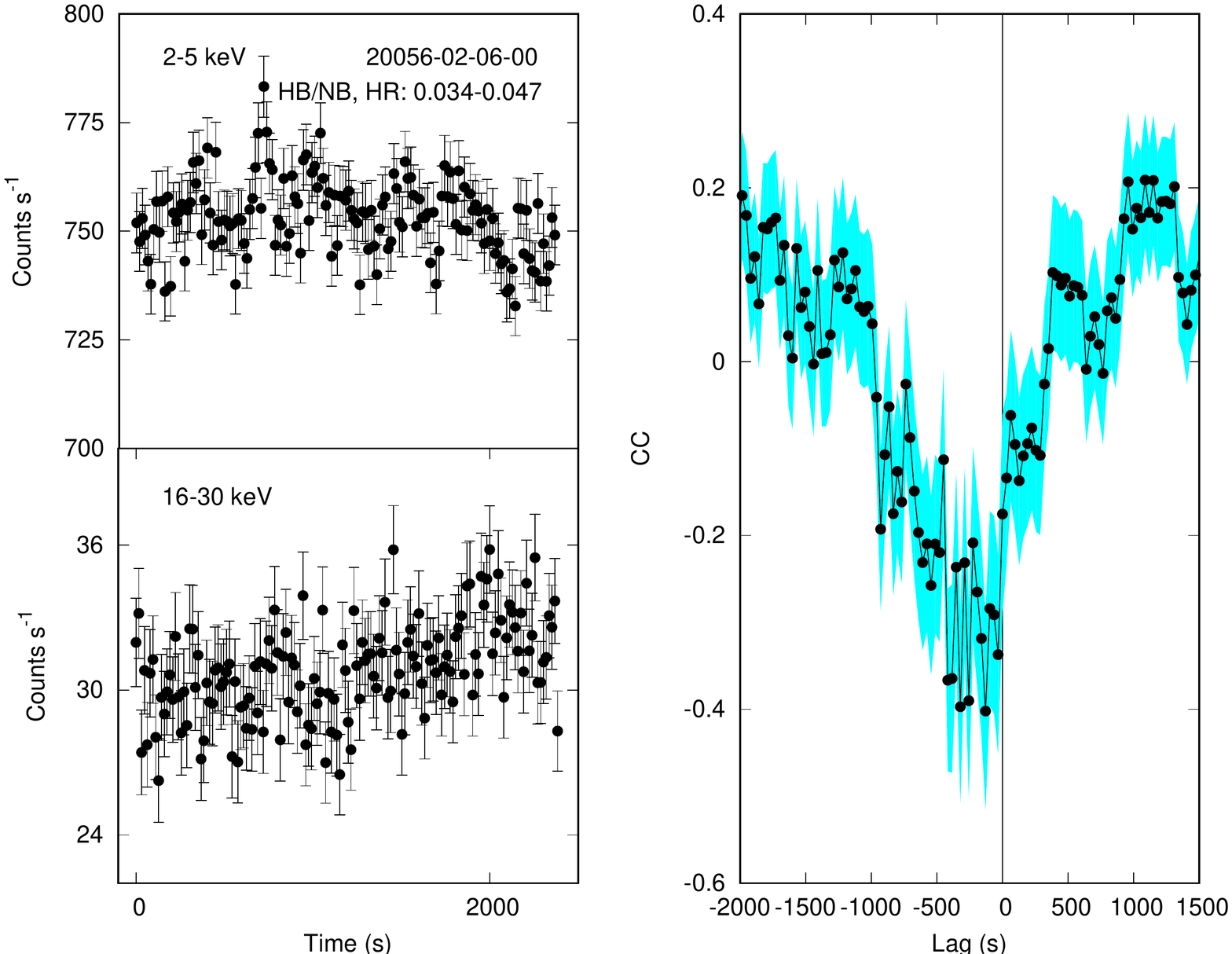}\\
\caption{Same as Figure 3a.}
\end{figure}
\begin{figure}
\includegraphics[width=0.25\paperwidth,angle=270]{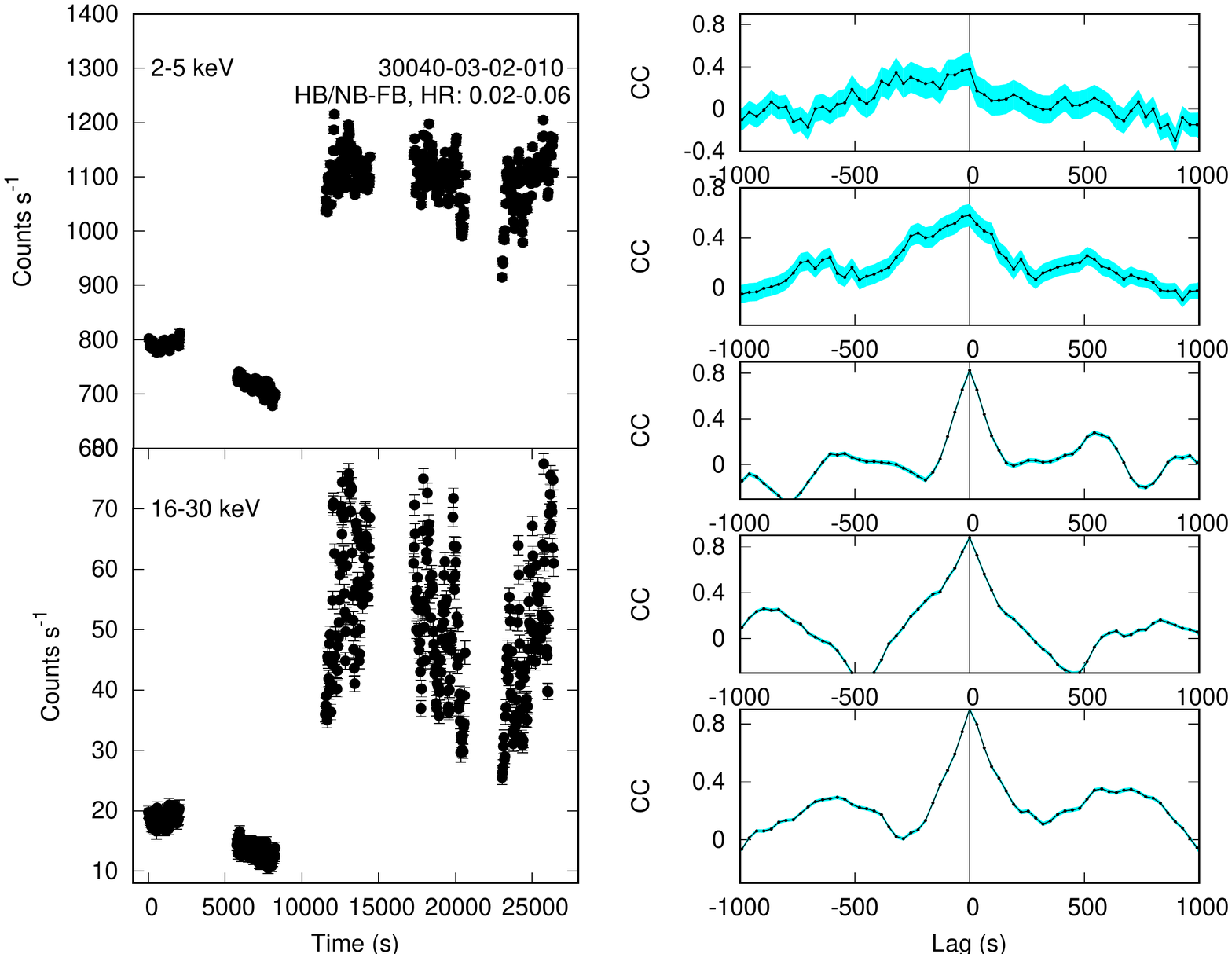} \\
\caption{Same as Figure 3a.}
\end{figure}
\begin{figure}
\includegraphics[width=0.25\paperwidth,angle=270]{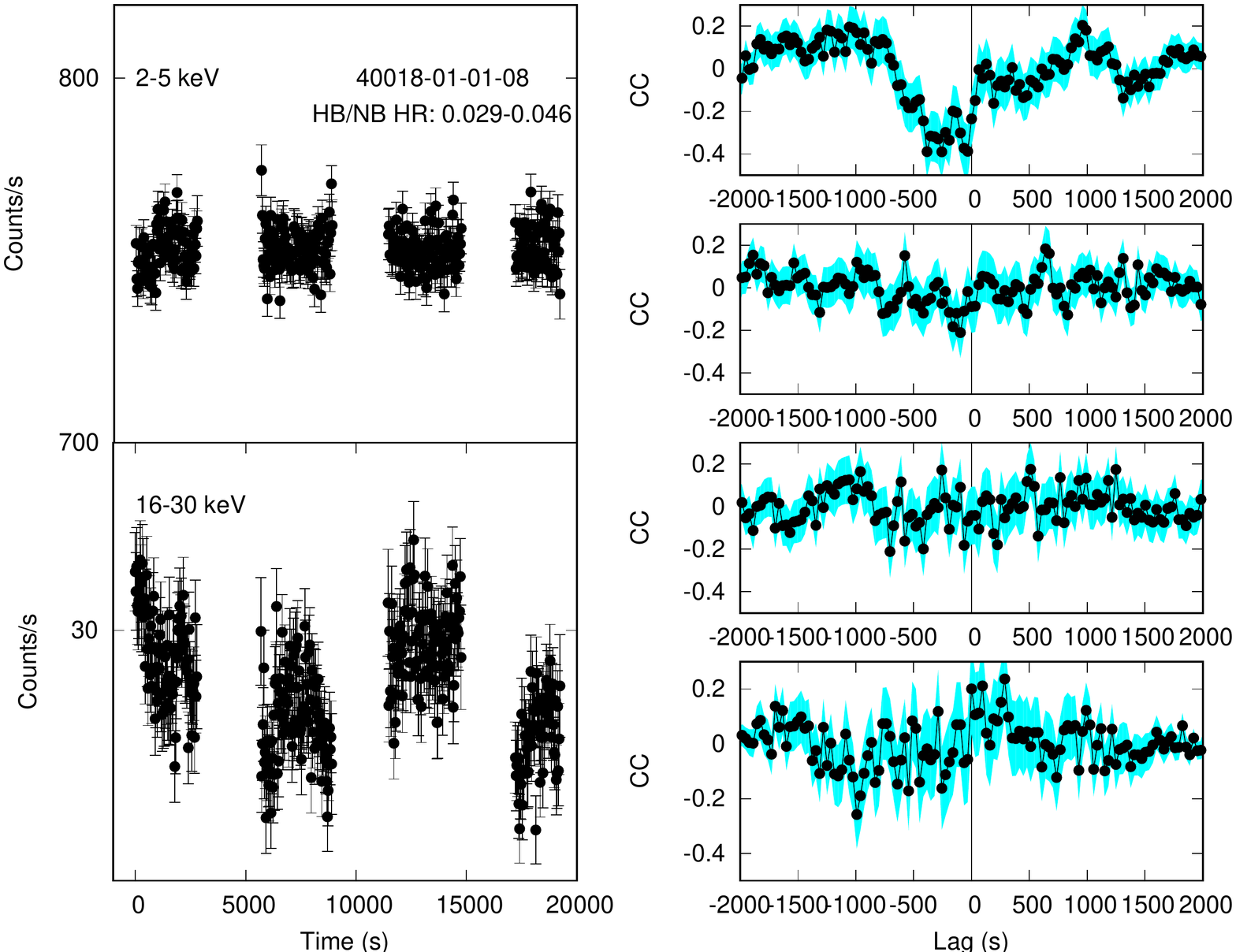}\\
\caption{Same as Figure 3a.}
\end{figure}
\begin{figure}
\includegraphics[width=0.25\paperwidth,angle=270]{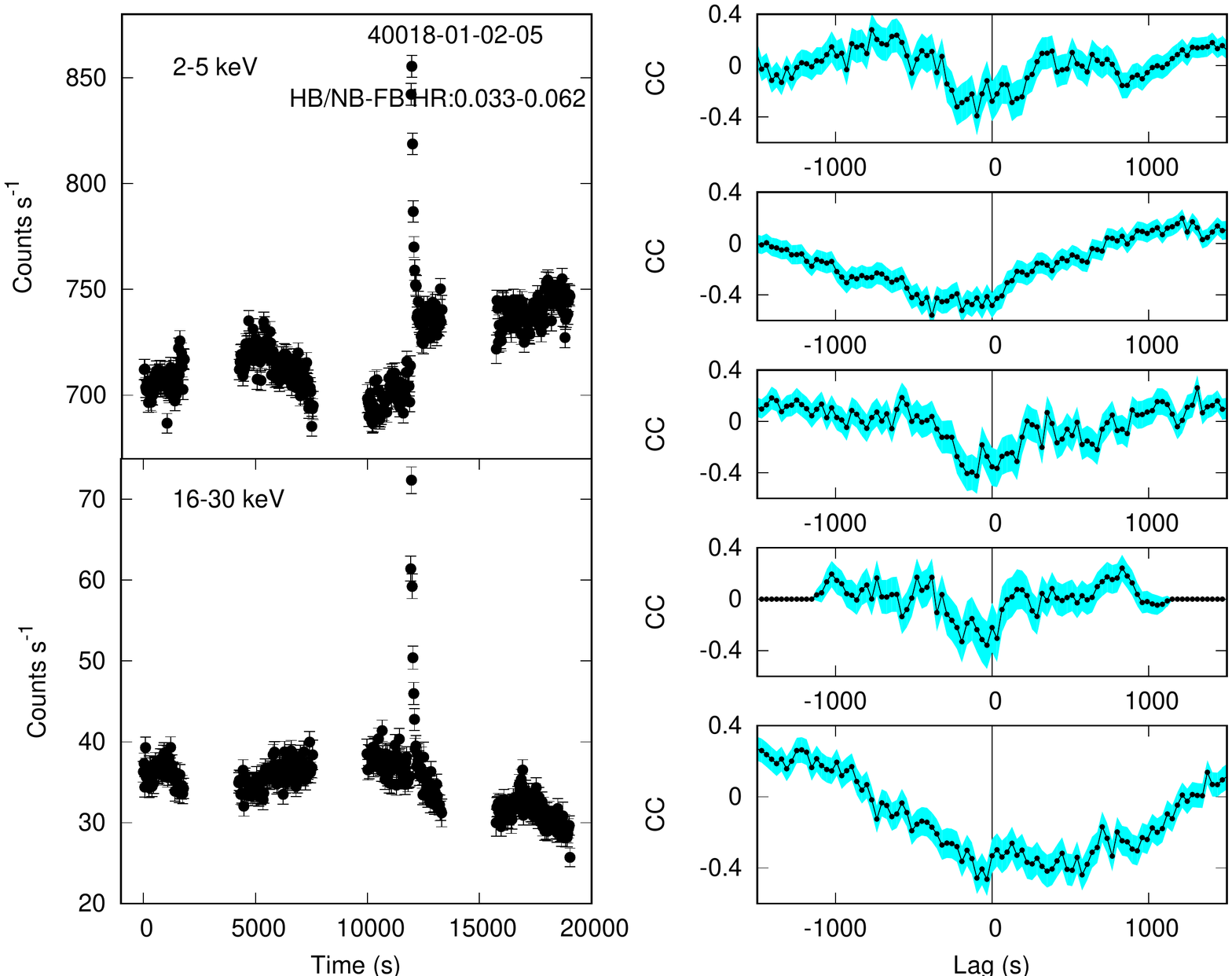}\\
\caption{Same as Figure 3a.}
\end{figure}
\begin{figure}
\includegraphics[width=0.25\paperwidth,angle=270]{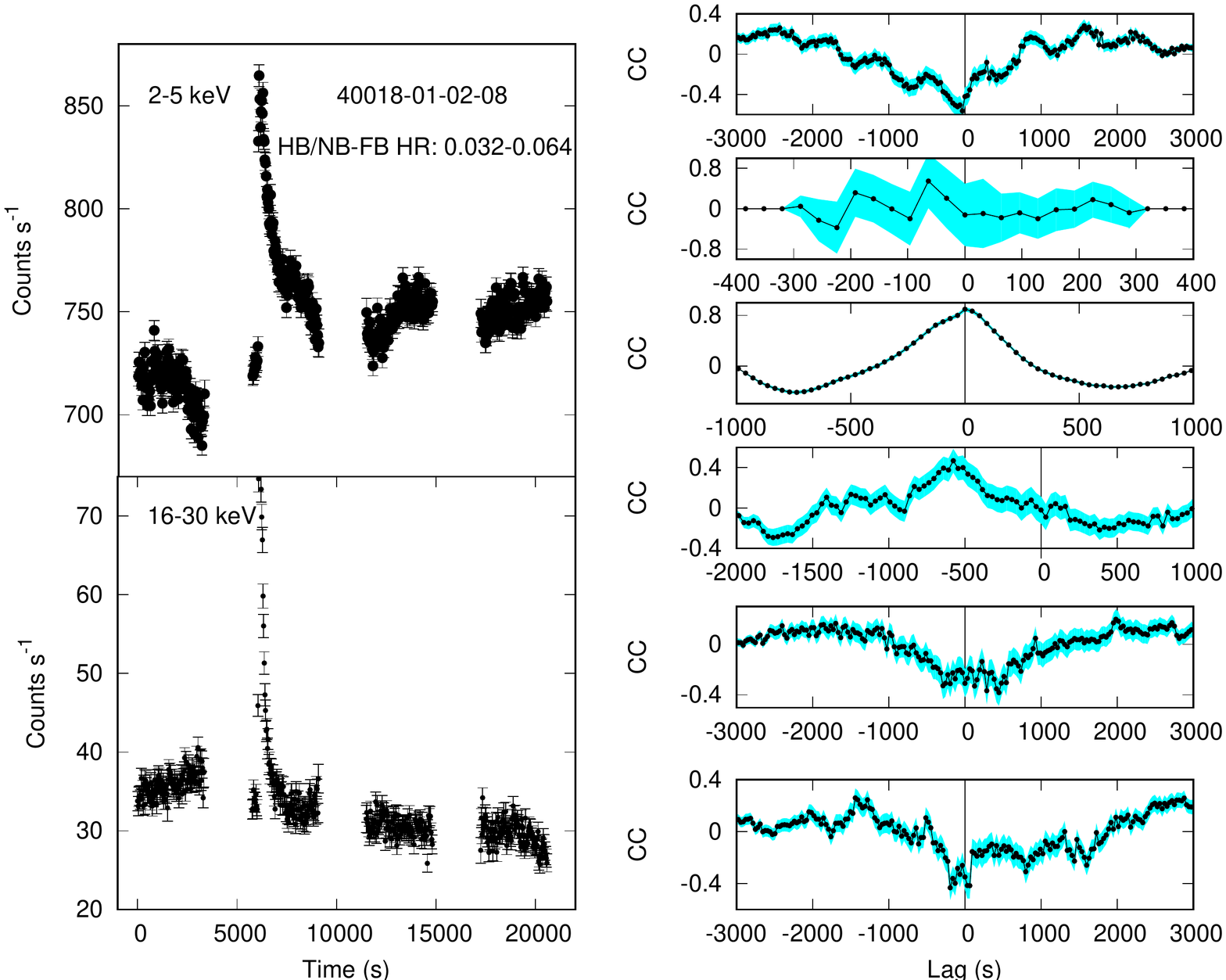}\\ \\ \\
\caption{Same as Figure 3a. Shaded regions of standard deviation of the CCFs are present although they are too small to be seen in the third panel.}
\end{figure}
\begin{figure}
\includegraphics[width=0.25\paperwidth,angle=270]{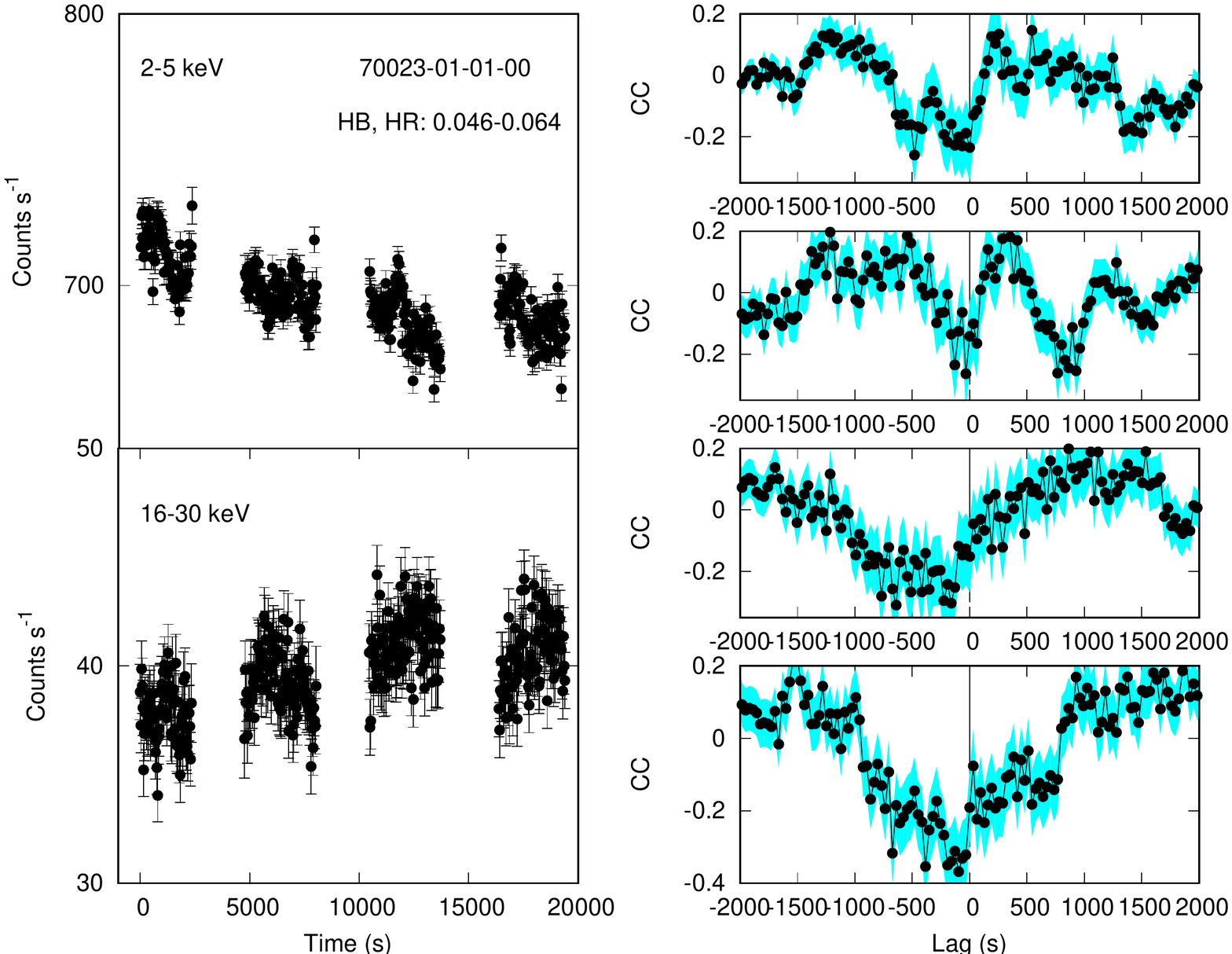}\\
\caption{Same as Figure 3a.}
\end{figure}
\begin{figure}
\includegraphics[width=0.25\paperwidth,angle=270]{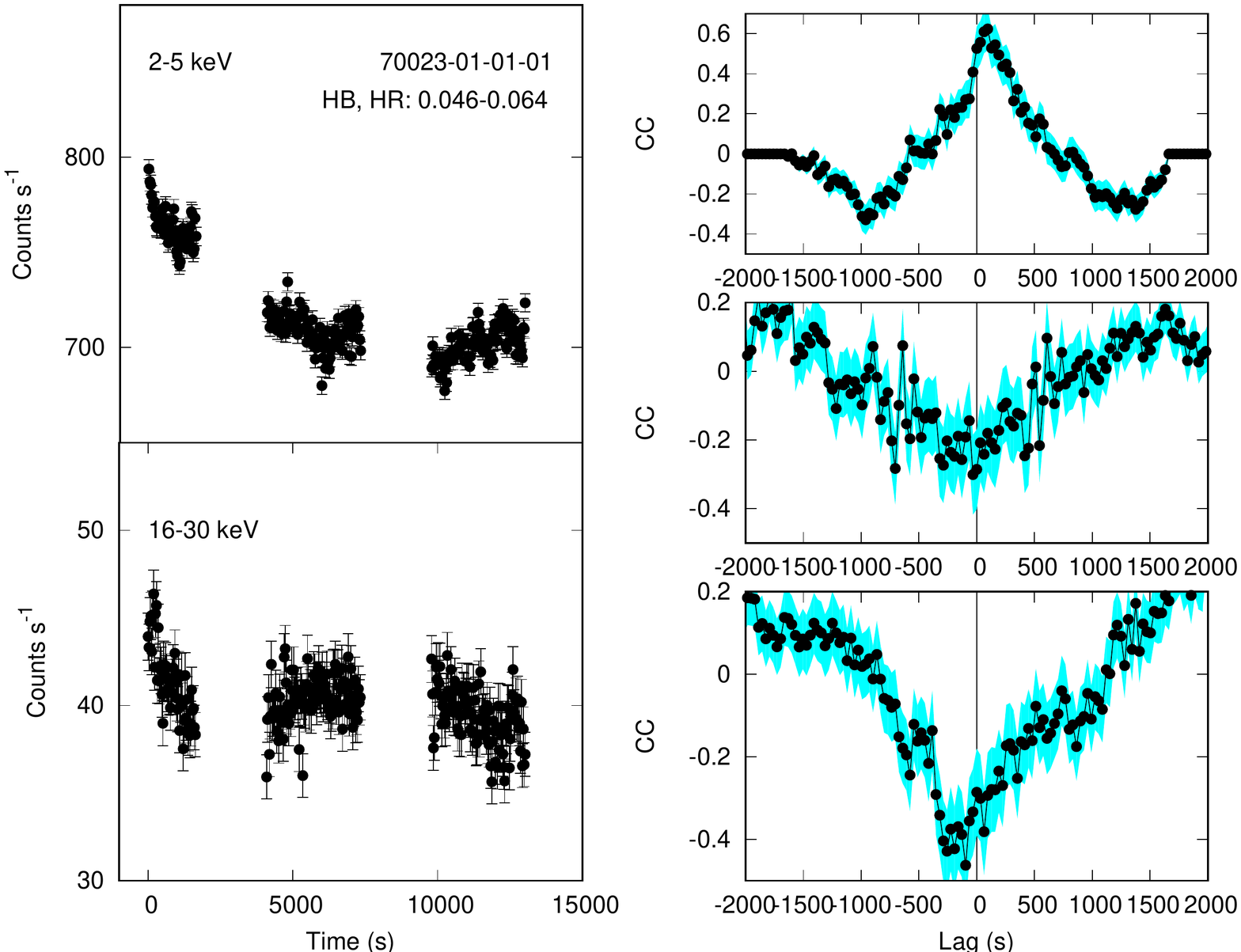}\\
\caption{Same as Figure 3a.}
\end{figure}
\begin{figure}
\includegraphics[width=0.25\paperwidth,angle=270]{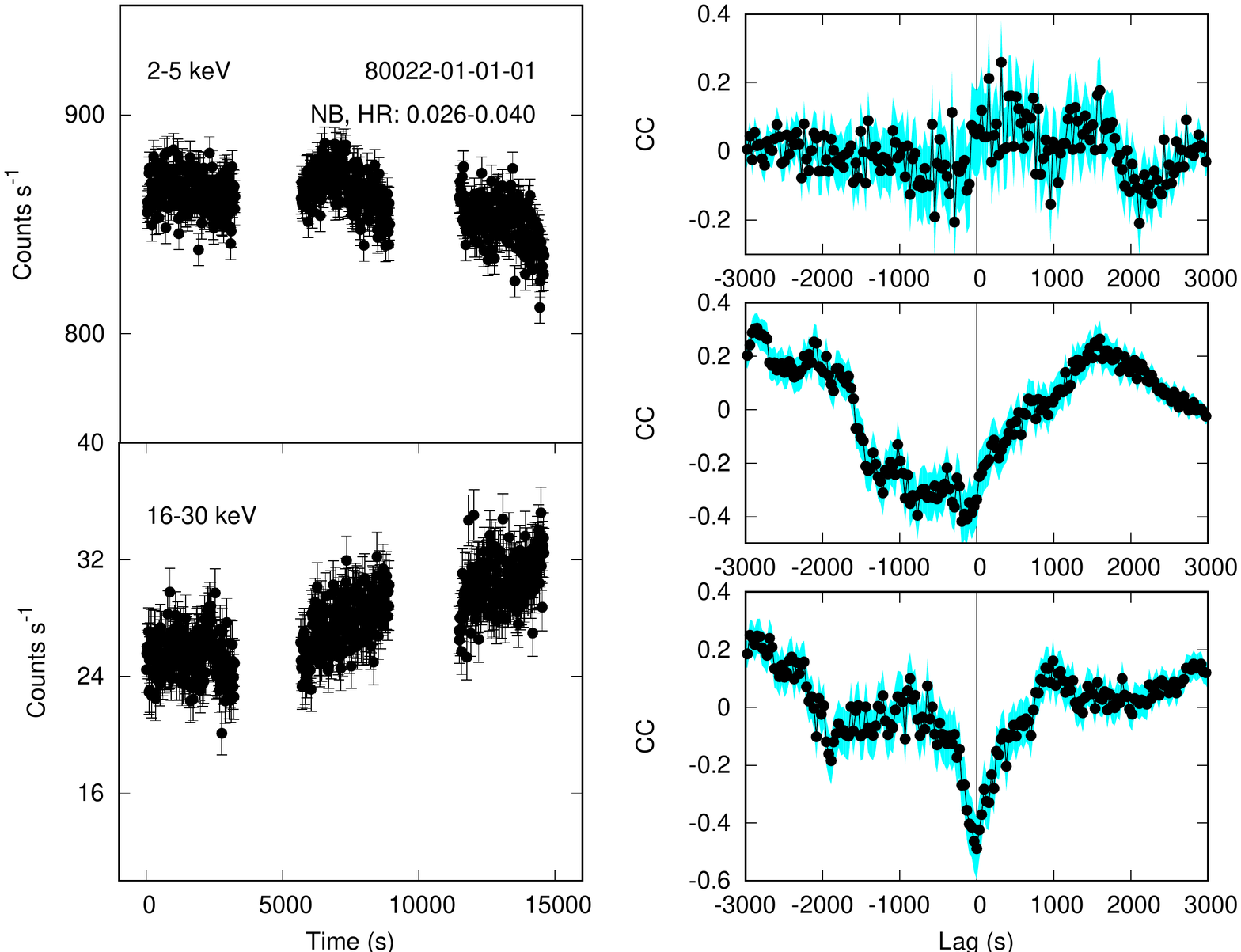}\\
\caption{Same as Figure 3a.}
\end{figure}
\begin{figure}
\includegraphics[width=0.25\paperwidth,angle=270]{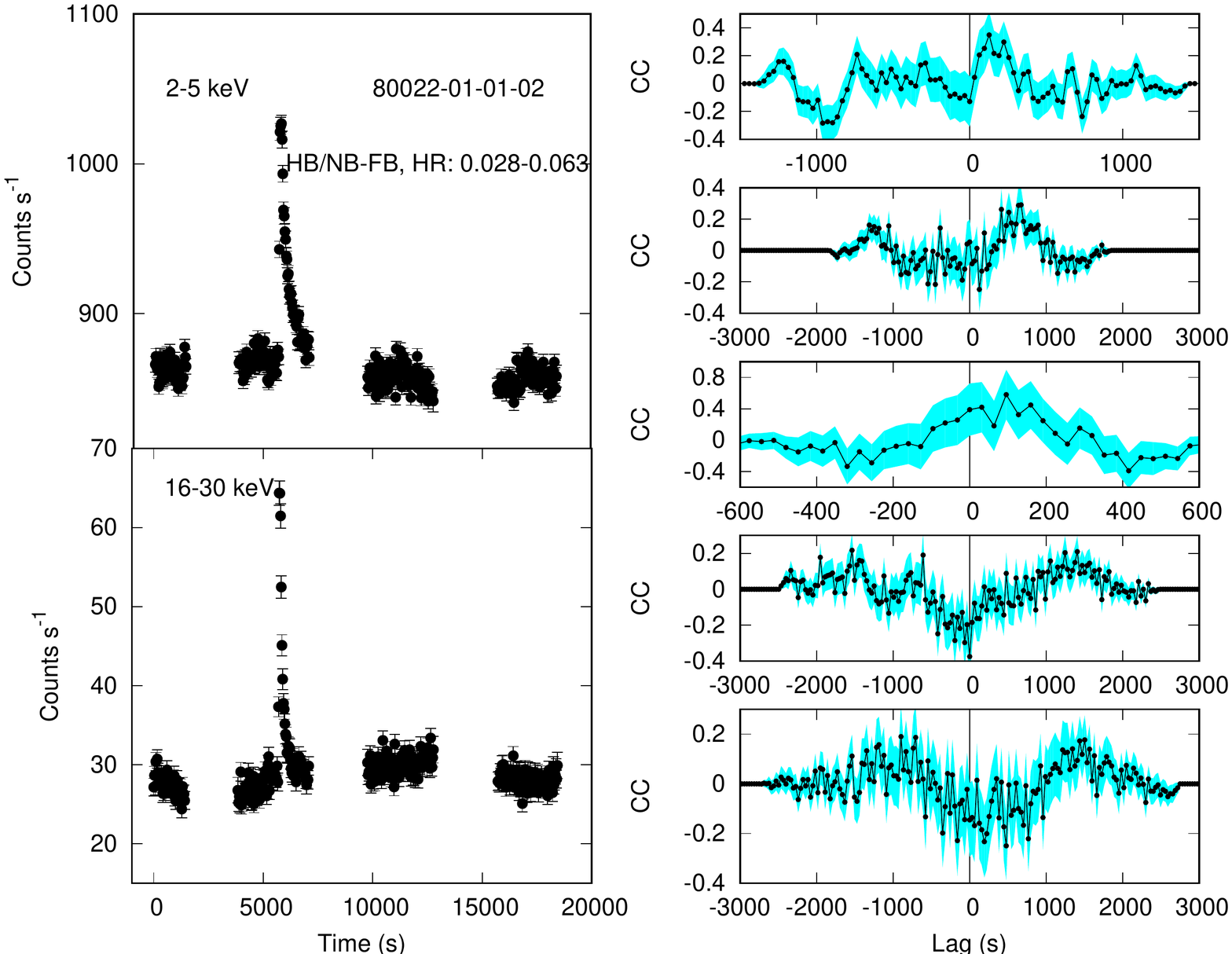}\\
\caption{Same as Figure 3a.}
\end{figure}
\begin{figure}
\includegraphics[width=0.25\paperwidth,angle=270]{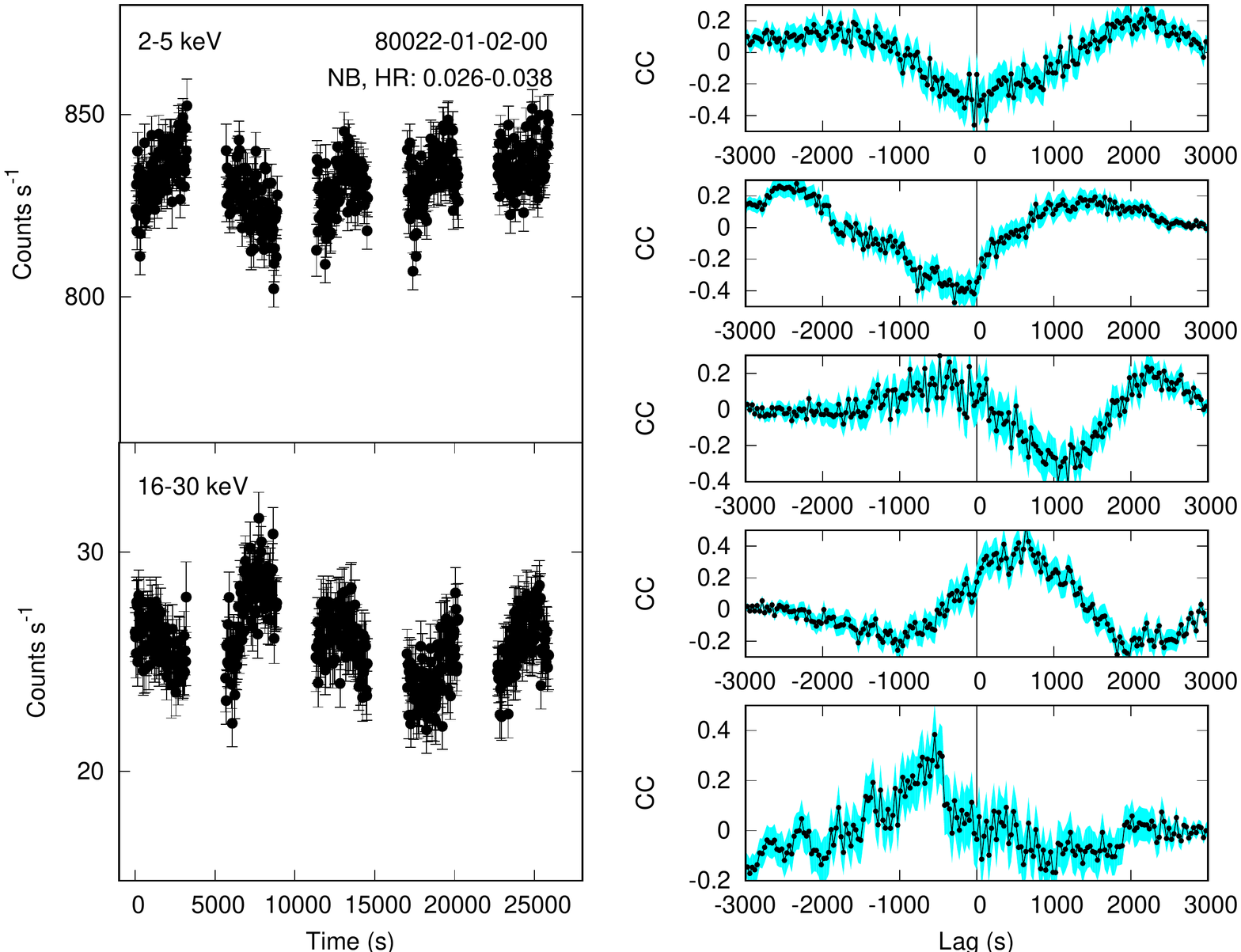}\\
\caption{Same as Figure 3a.}
\end{figure}
\begin{figure}
\includegraphics[width=0.25\paperwidth,angle=270]{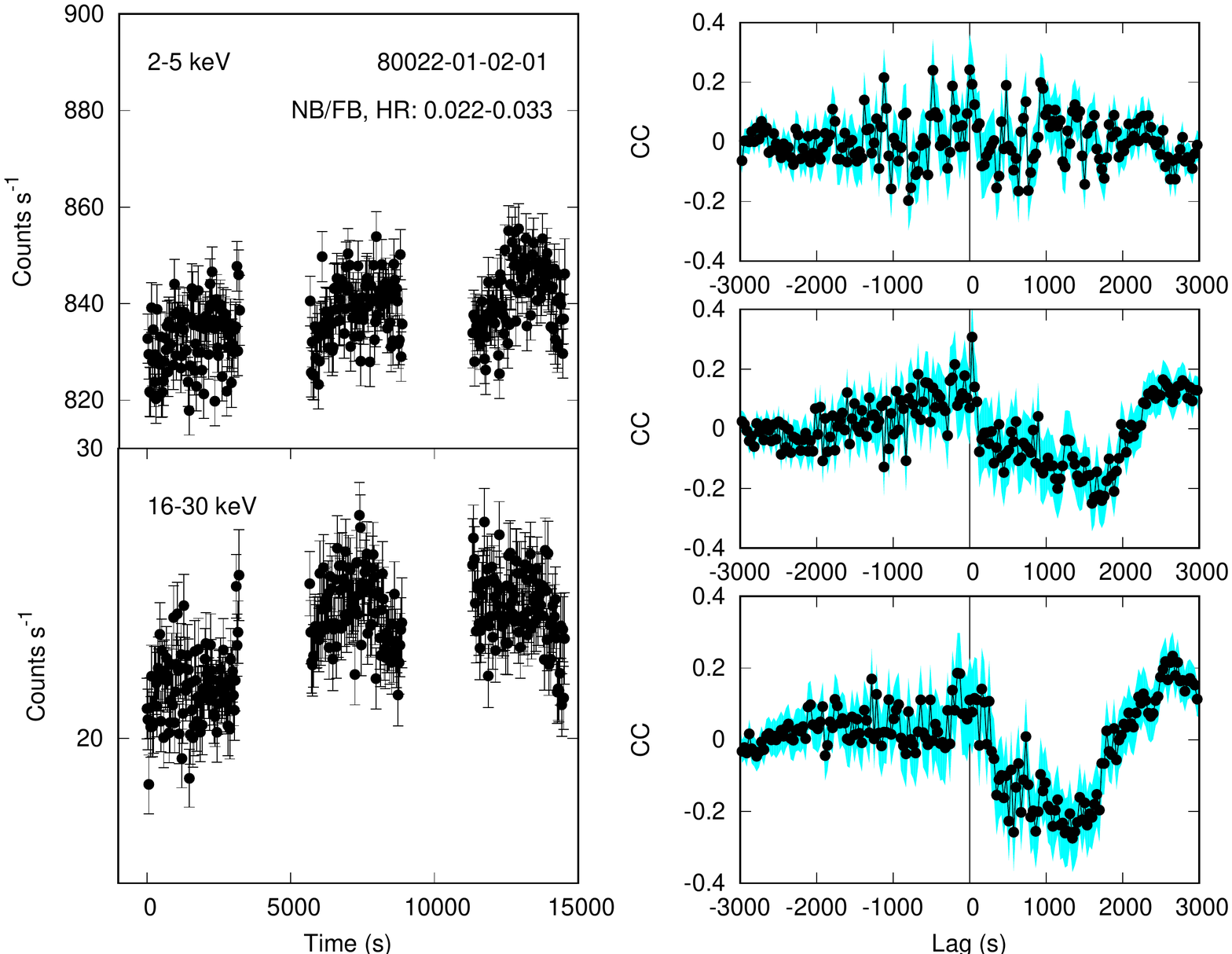}\\
\caption{Same as Figure 3a.}
\end{figure}
\begin{figure}
\includegraphics[width=0.25\paperwidth,angle=270]{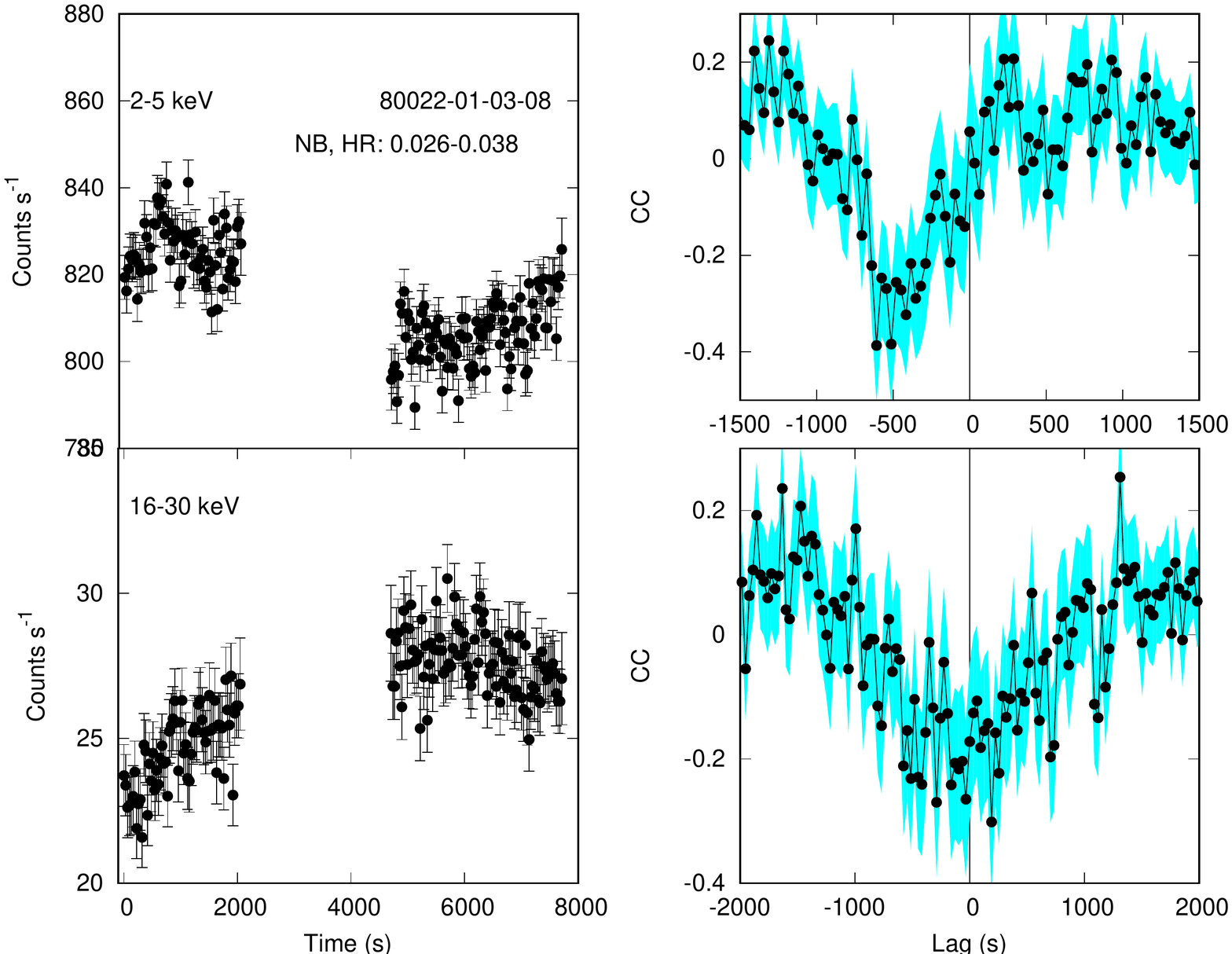}\\
\caption{Same as Figure 3a.}
\end{figure}
\begin{figure}
\includegraphics[width=0.25\paperwidth,angle=270]{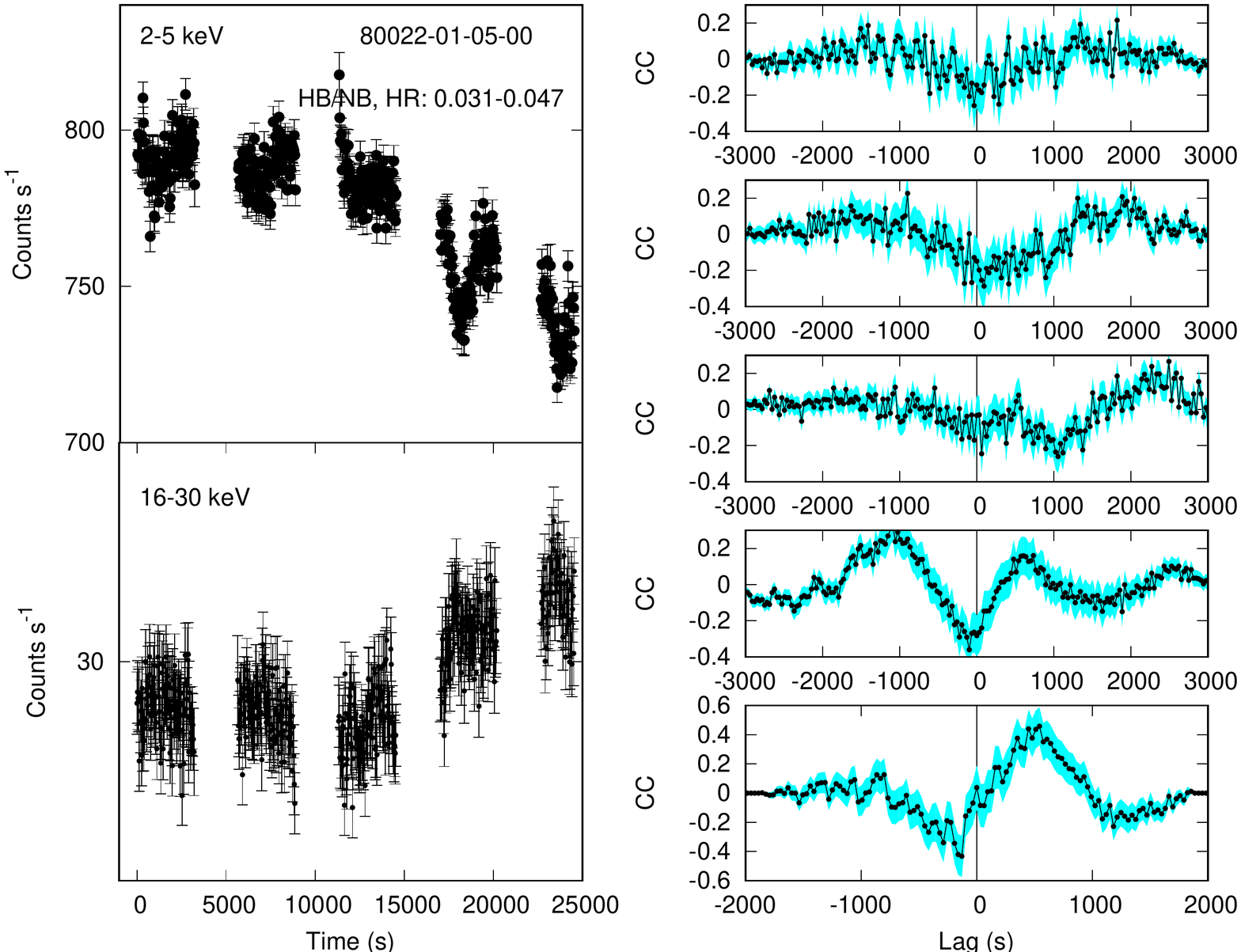}\\
\caption{Same as Figure 3a.}
\end{figure}
\begin{figure}
\includegraphics[width=0.25\paperwidth,angle=270]{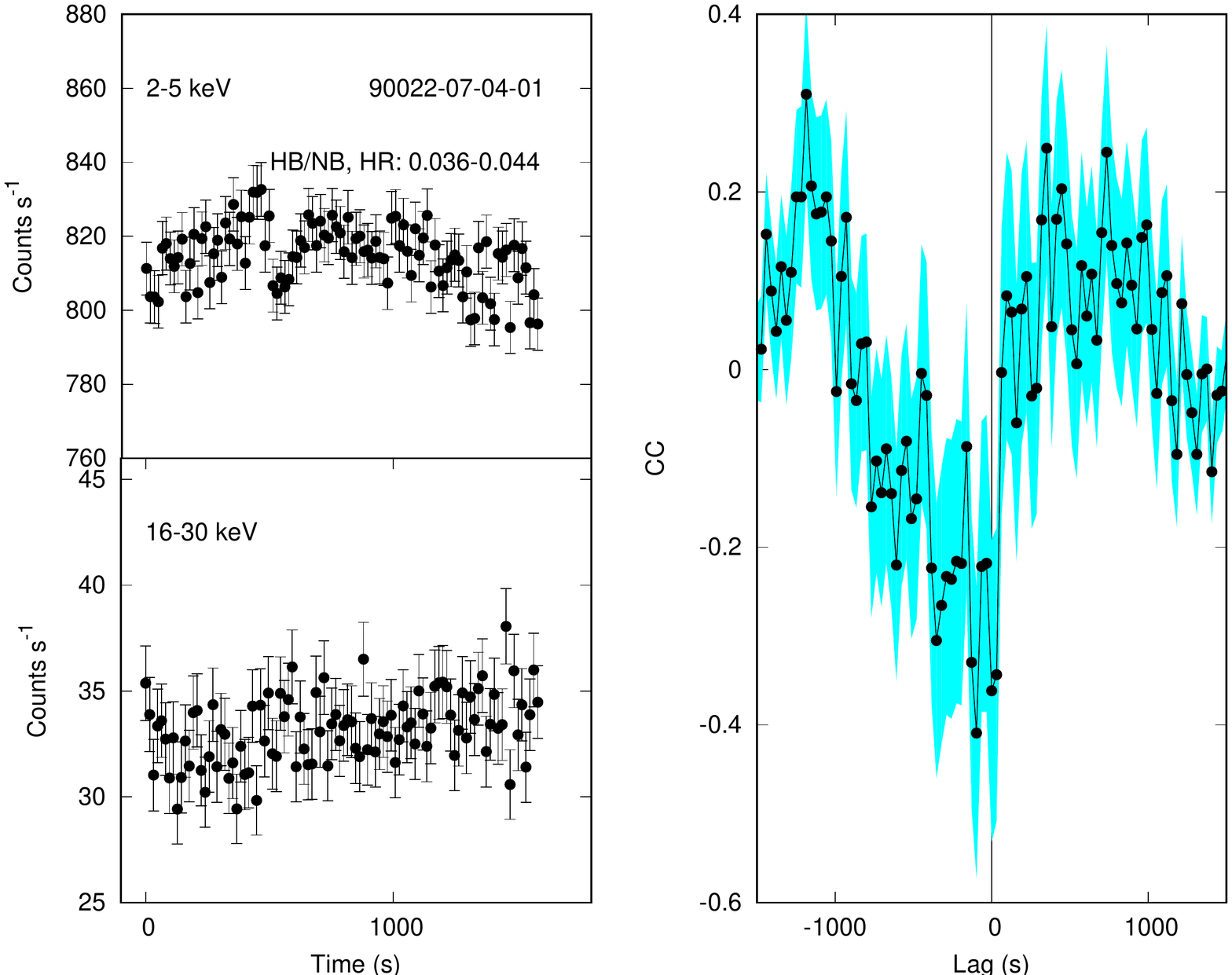}\\
\caption{Same as Figure 3a.}
\end{figure}
\begin{figure}
\includegraphics[width=0.25\paperwidth,angle=270]{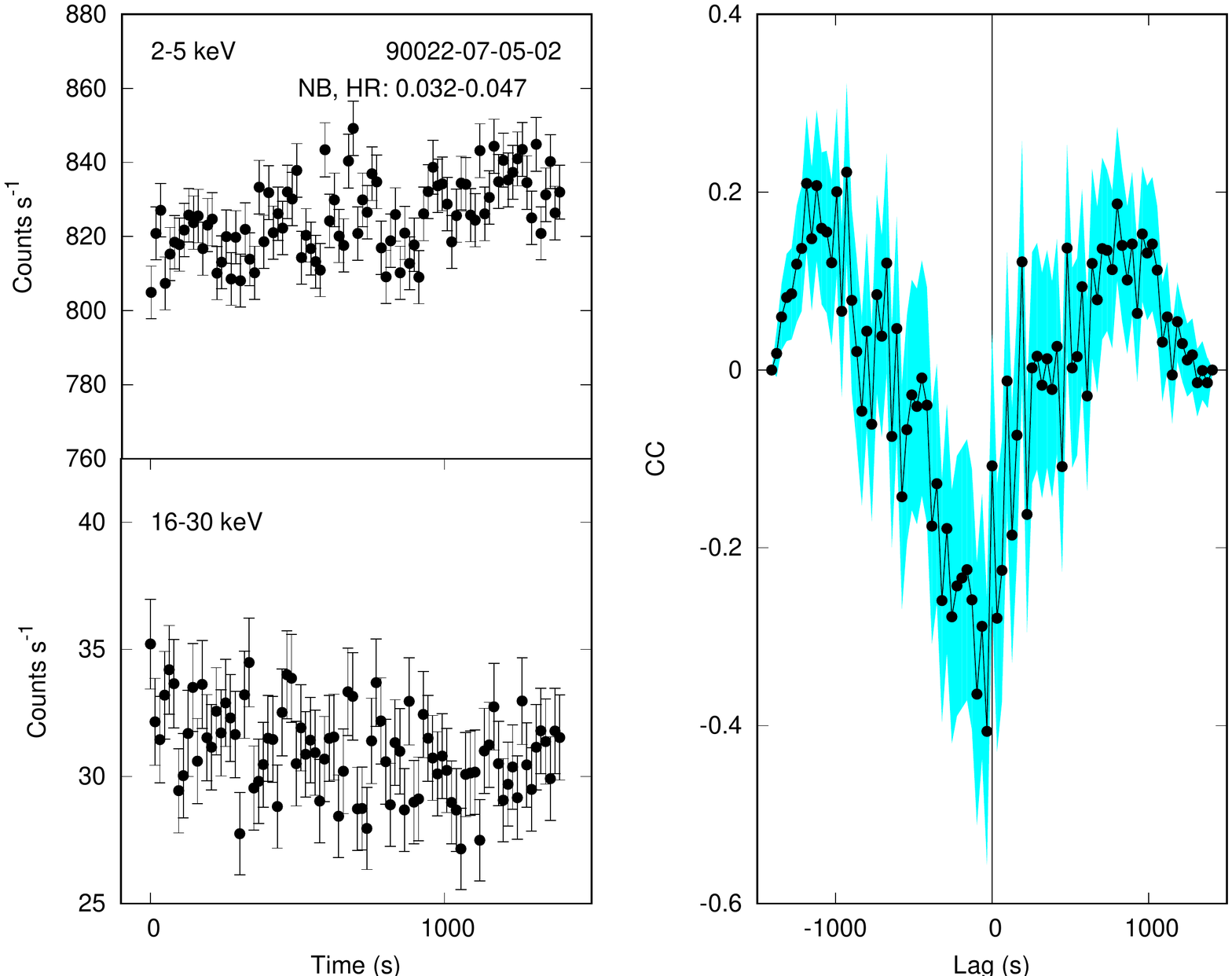}\\
\caption{Same as Figure 3a.}
\end{figure}
\begin{figure}
\includegraphics[width=0.25\paperwidth,angle=270]{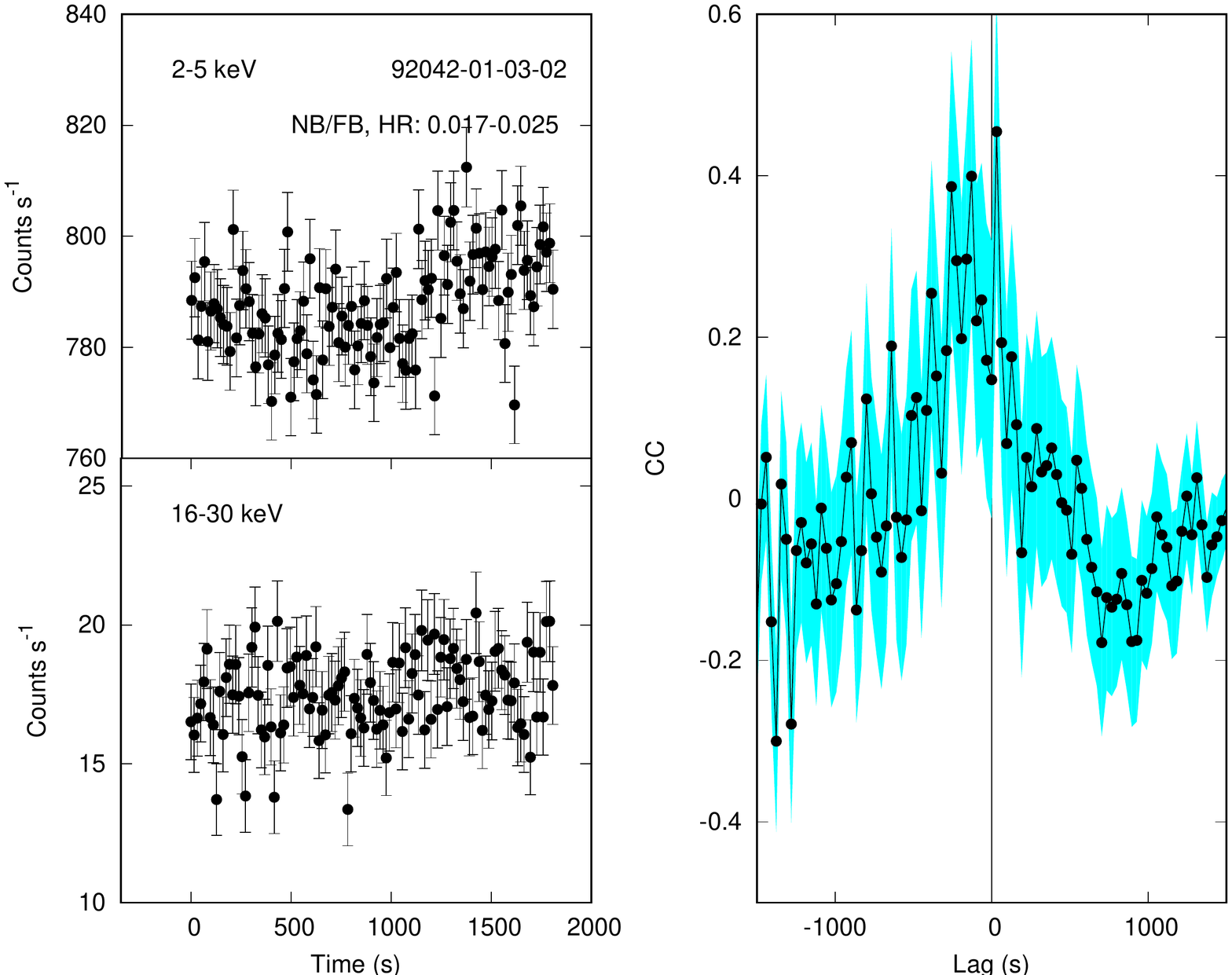}\\
\caption{Same as Figure 3a.}
\end{figure}
\begin{figure}
\includegraphics[width=0.25\paperwidth,angle=270]{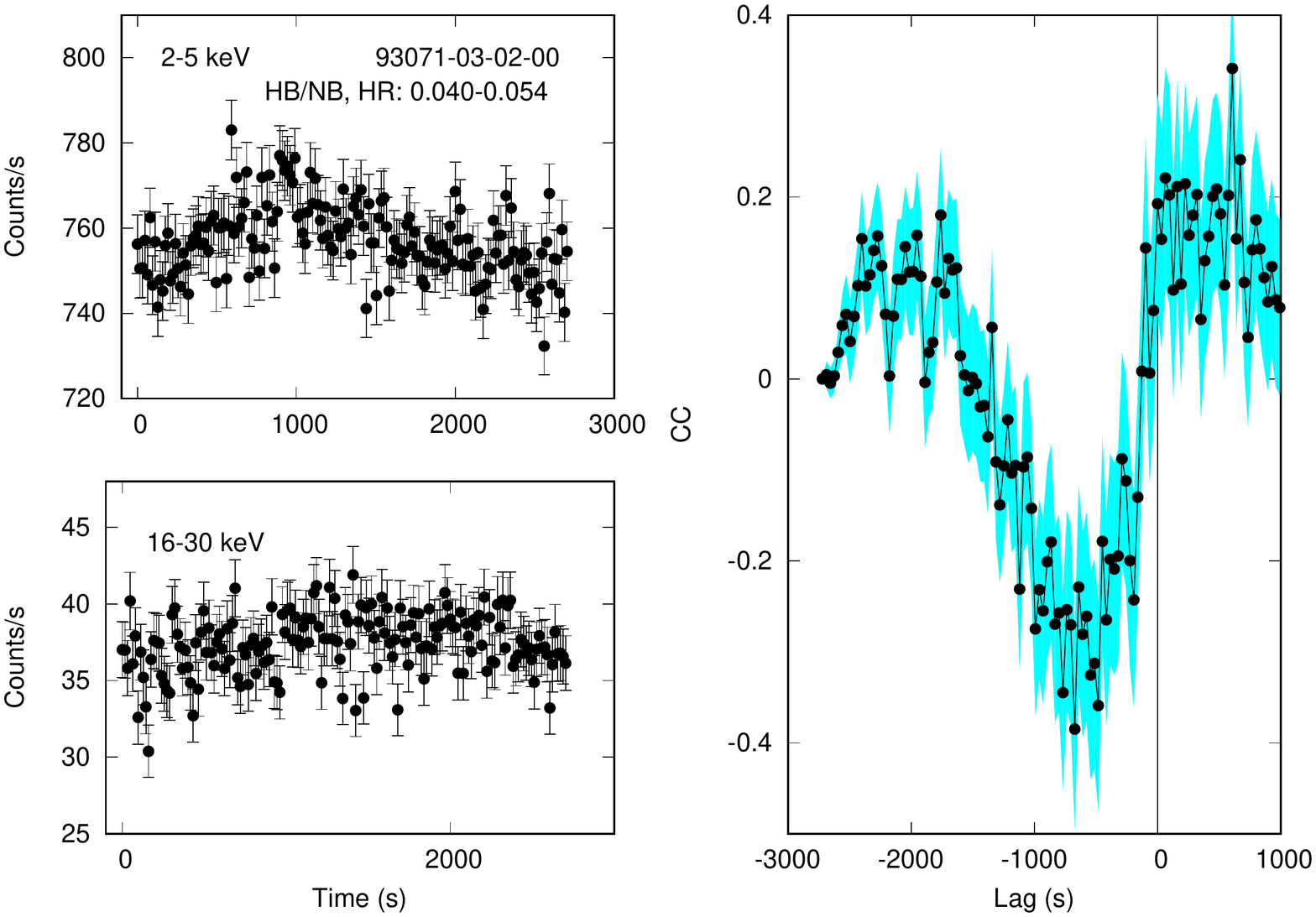}\\
\caption{Same as Figure 3a.}
\end{figure}
\begin{figure}
\includegraphics[width=0.25\paperwidth,angle=270]{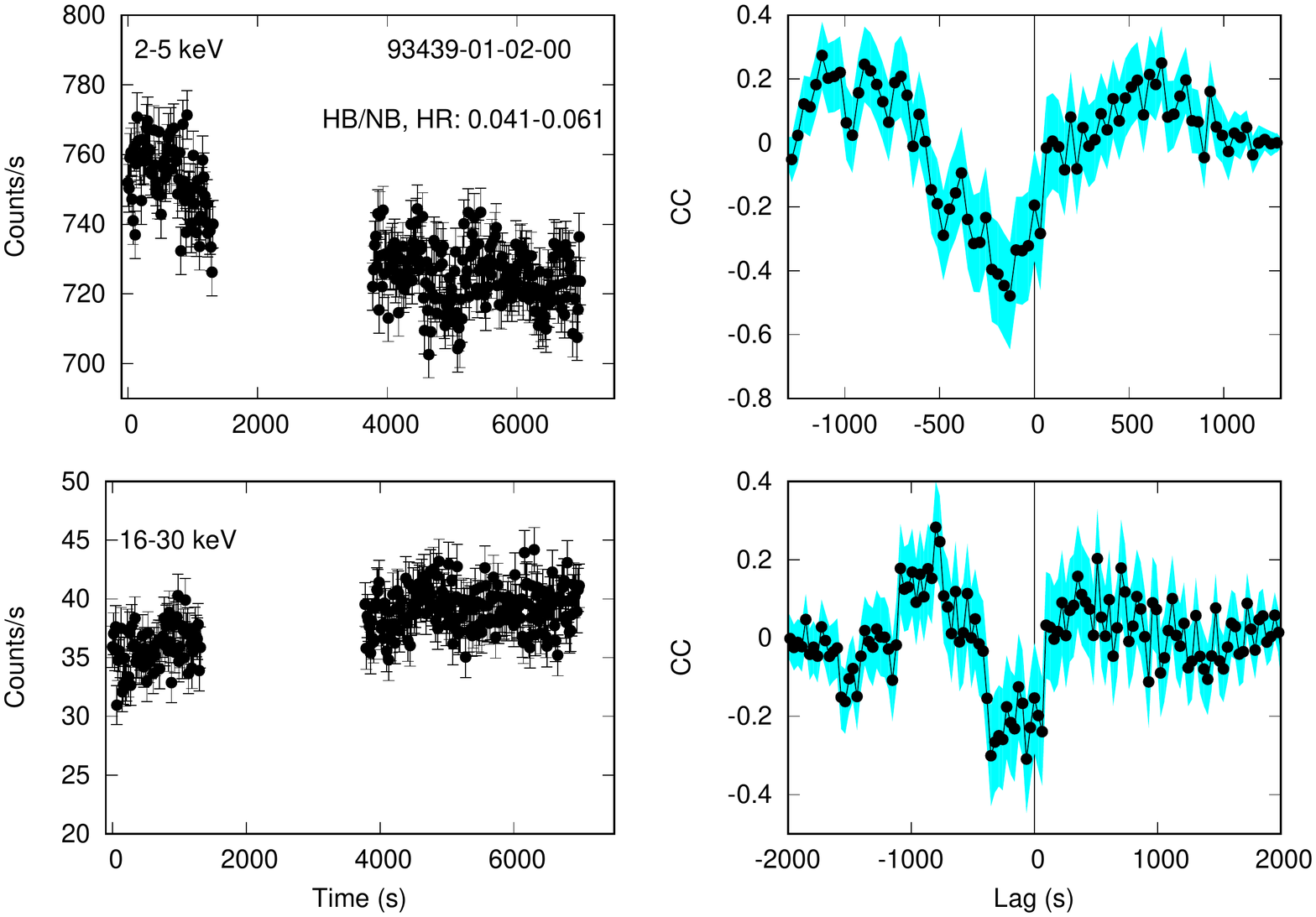}\\
\caption{Same as Figure 3a.}
\end{figure}


\end{subfigures}
\begin{figure*}

\includegraphics[width=8.0cm,height=18 cm,angle=270]{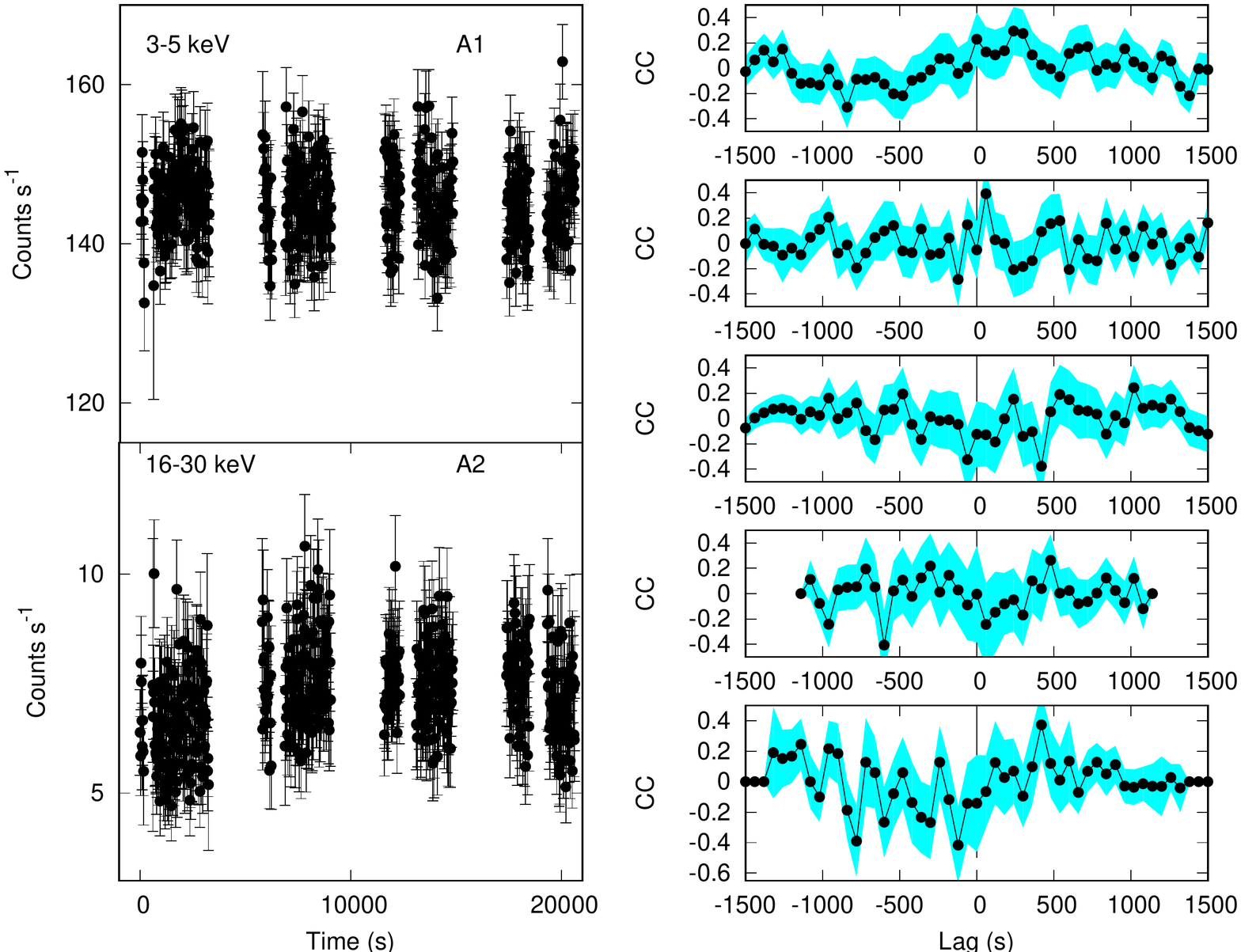}\\ \\ \\
\includegraphics[width=8.0cm,height=18 cm,angle=270]{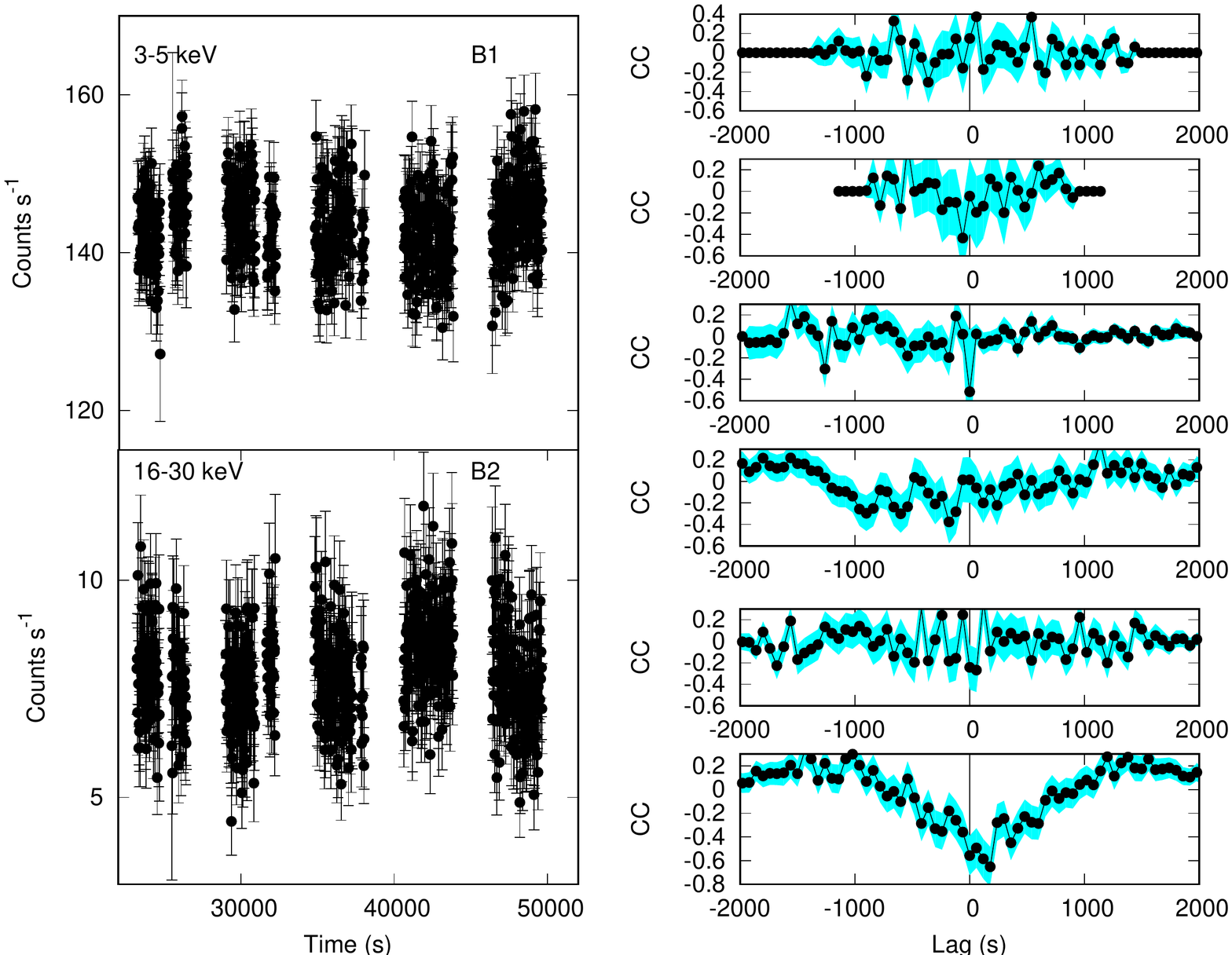}\\ \\ \\
\caption{The background subtracted 20 s bin NuSTAR  soft (3--5 keV) and hard X-ray (16--30 keV) light curves (left panels) for which CCF lags are shown (right panels). Light curve has been split first into 5 major sections in the top two panels (A1 (3-5 keV), A2 (16-30 keV) ) and next into 6 major sections in the bottom two panels (B1 (3-5 keV), B2 (16-30 keV) ) to effectively display the CCFs. The energy bands are mentioned in the light curves. Shaded regions in the right panels show the standard deviation of the CCF.}

\end{figure*}

\begin{figure*}

\includegraphics[width=10.0cm,height=18 cm,angle=270]{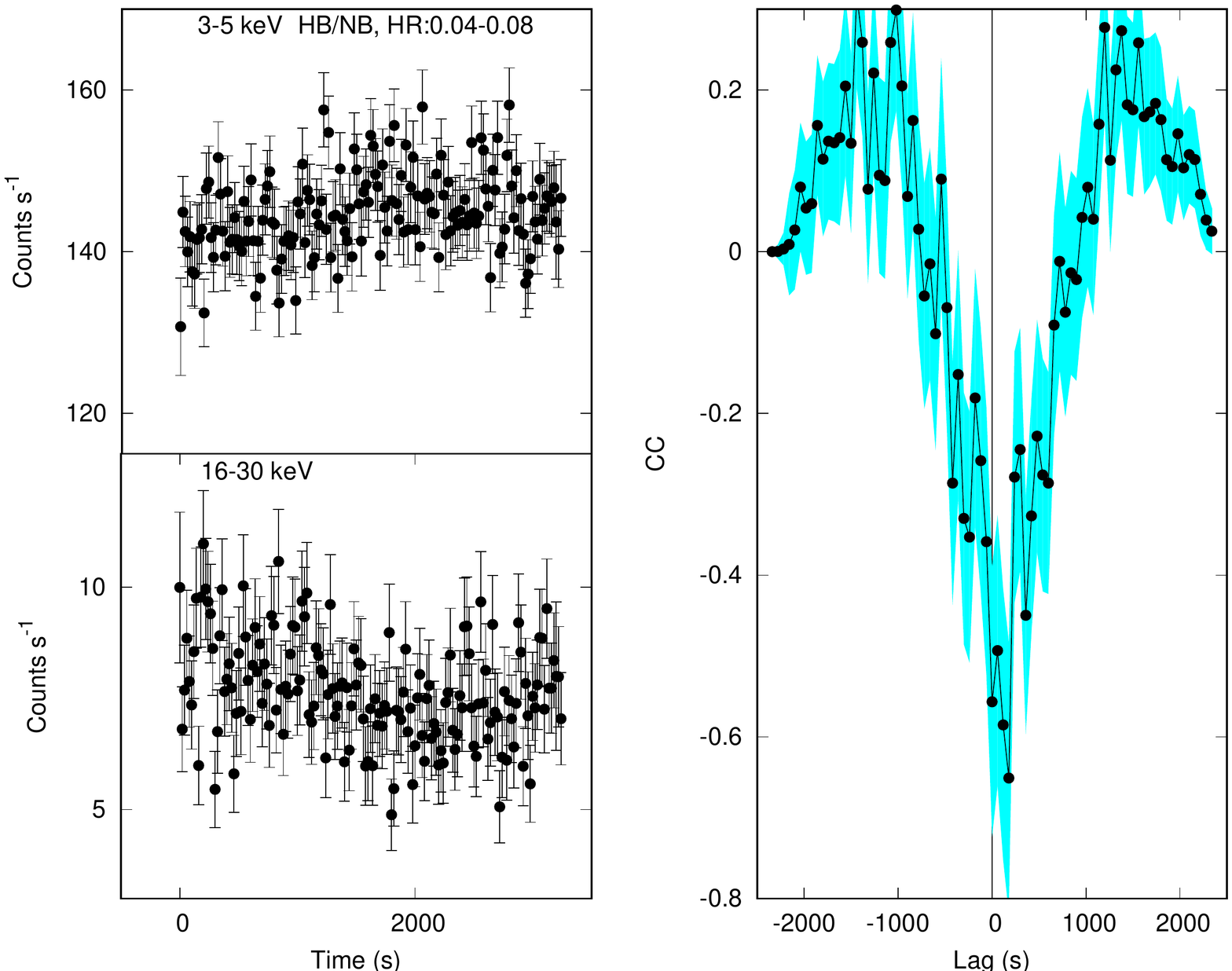}\\ \\ \\
\caption{The last section of Figure 3 is replotted for clarity i.e. background subtracted 20 s bin NuSTAR soft (3--5 keV) and hard X-ray (16--30 keV) light curves (left panels) for which CCF lags are shown (right panel). Vertical line shows the zero lag and a hard X-ray delay is observed.}

\end{figure*}

\begin{figure*}
\includegraphics[width=10.0cm,height=15 cm,angle=270]{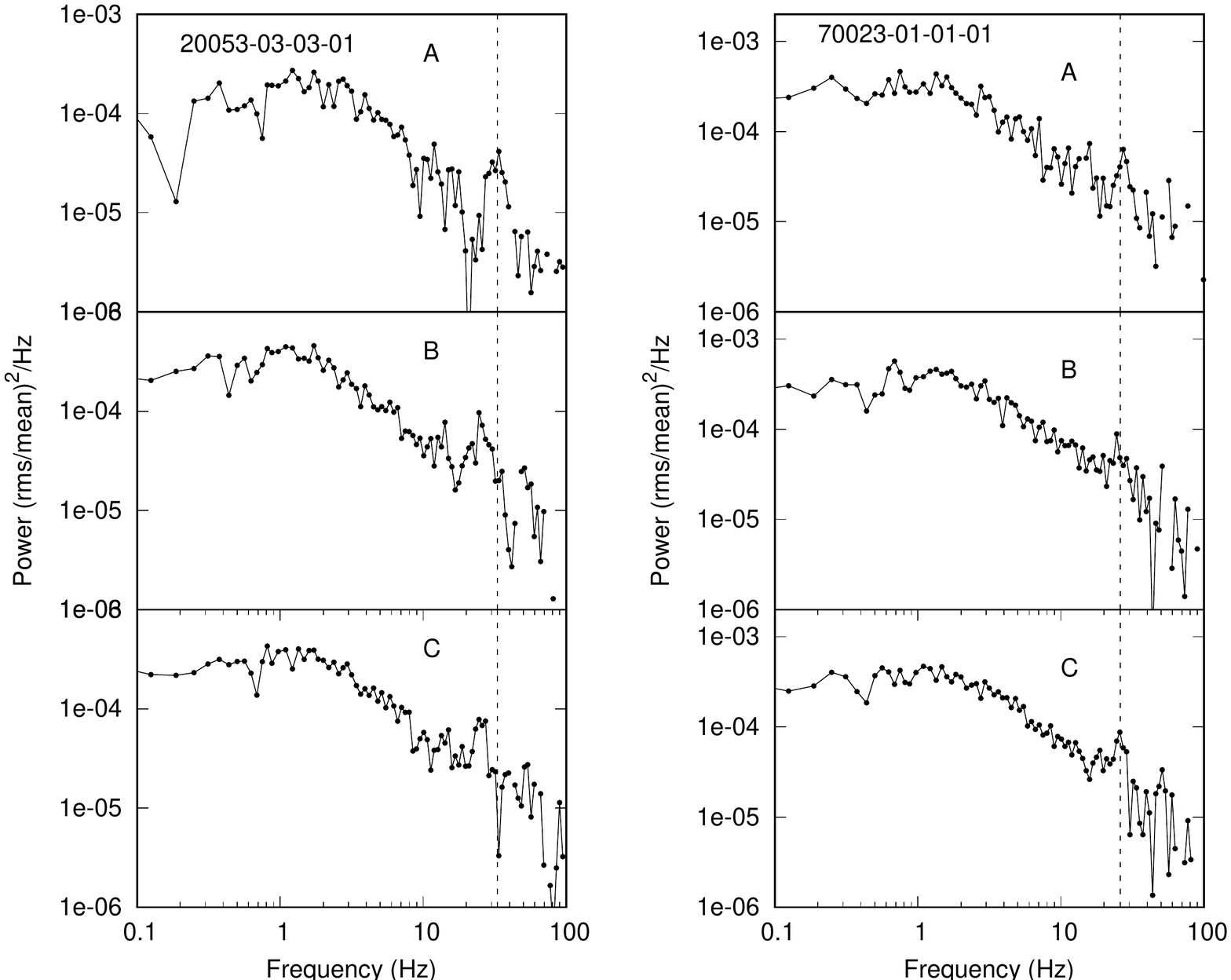}\\ \\ \\

\caption{The power density spectra of each section of  respective ObsIDs are shown. The dashed vertical lines in ObsID 20053-03-03-01 (left panels) and ObsID 70023-01-01-01 (right panels) are at 33 Hz and 26 Hz QPO.}
\end{figure*}







\begin{figure*}
\raggedleft

\includegraphics[width=8.0cm,height=15.0cm,angle=270]{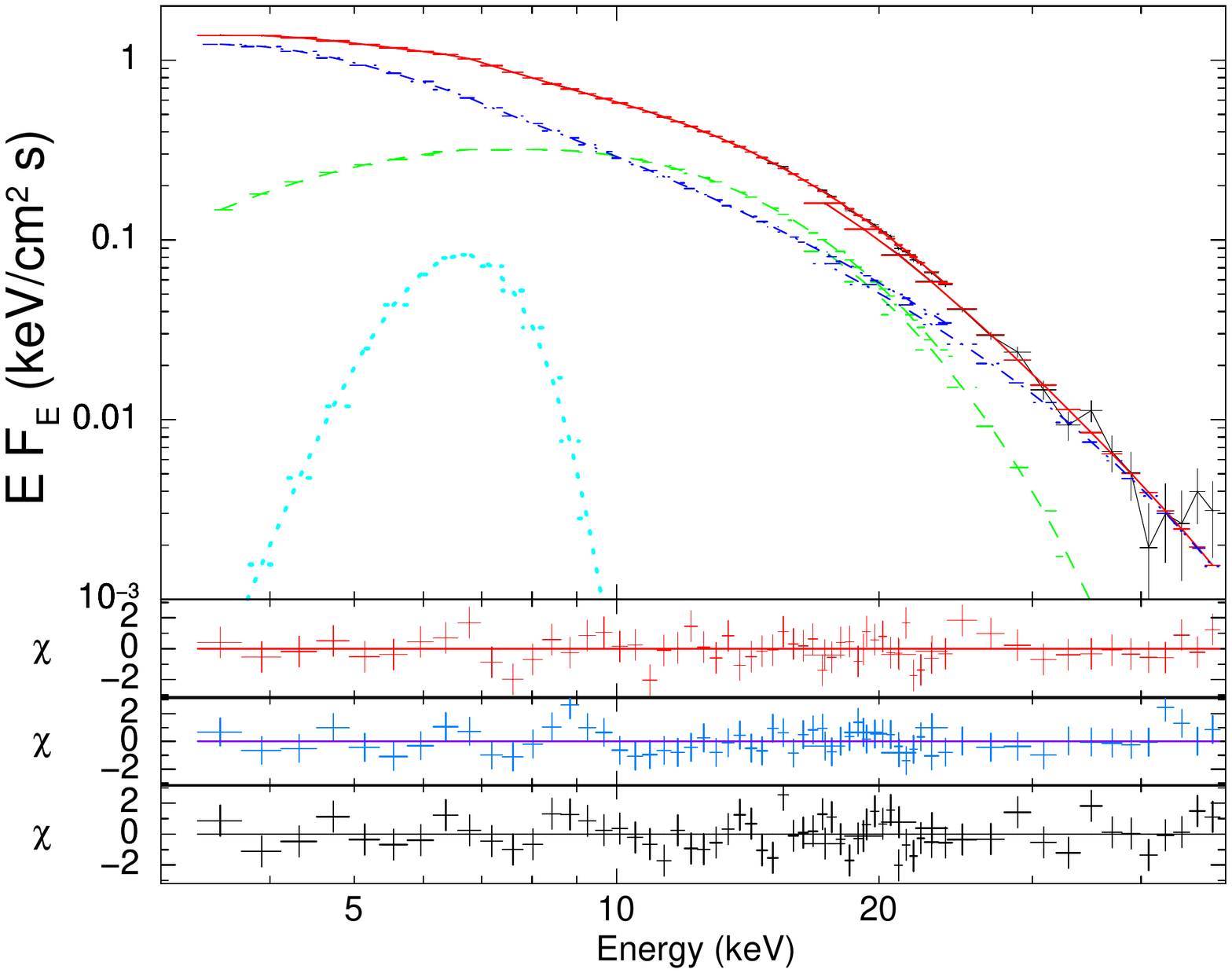} \\
\includegraphics[width=8.0cm, height=15.0cm,angle=270]{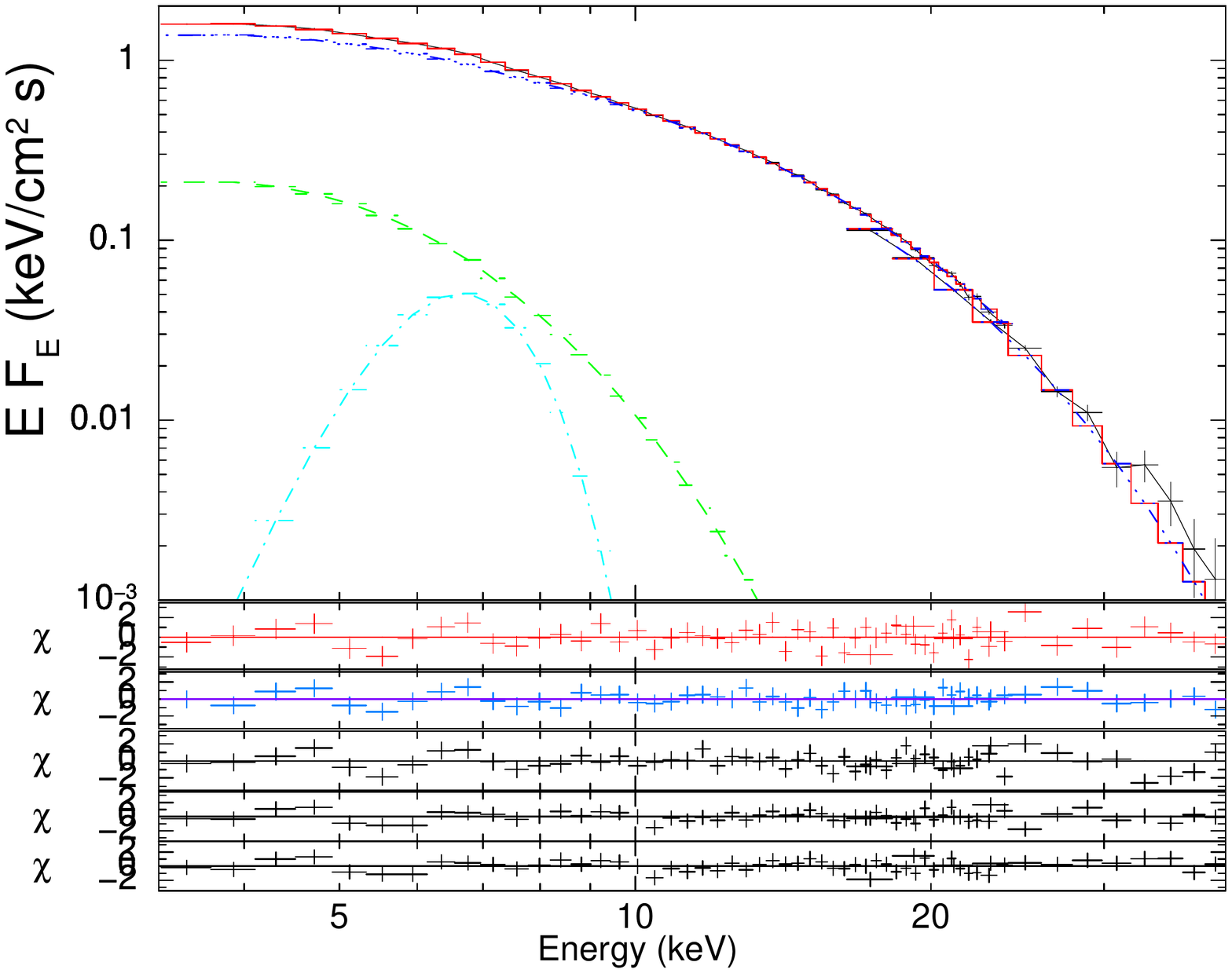}\\

\caption{ Top: The unfolded energy spectra in 2-50 keV band obtained from the best-fit spectral result of the RXTE PCA and HEXTE data of ObsID 70023-01-01-01 along with their model components (BB+Gaussian+nthcomp) and residuals. Bottom: Same figure for 80022-01-03-05.}
\end{figure*}








\begin{figure*}

\includegraphics[height=17cm,width=10cm, angle=270=]{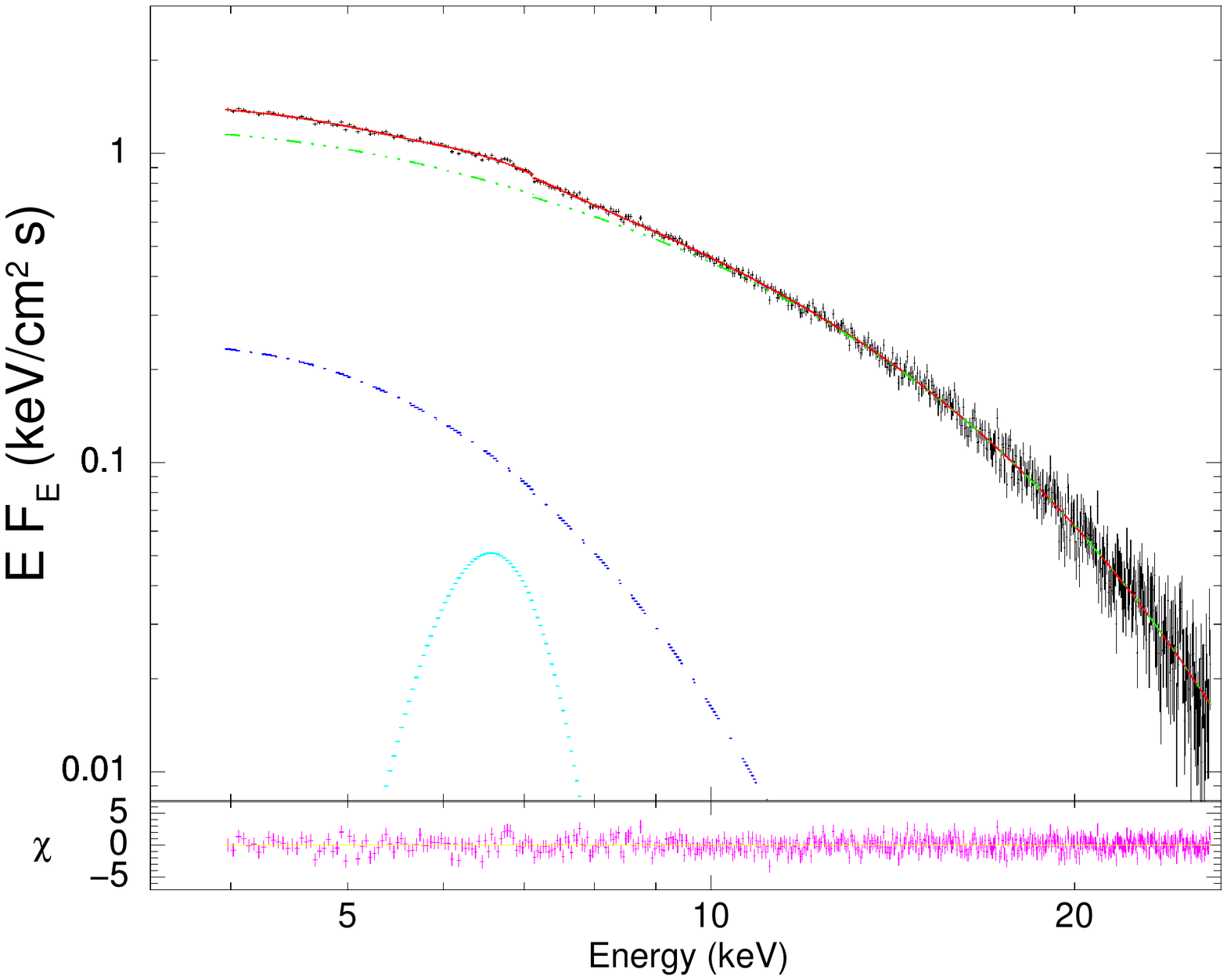}\\

\caption{The best fit {\it Nustar} spectrum unfolded with a model {\it wabs(bbody+Gaussian+nthcomp)} of the last section of light curve associated with a delay.  } 
\end{figure*}
\clearpage

\scriptsize
\longrotatetable
\begin{longtable*}{|l|l|l|l|l|l|l|l}
\caption{Cross correlation functions of GX 17+2 using 2-5 keV (soft) and 16-30 keV (hard) background subtracted 32 s bin RXTE PCA light curves. 
CC means cross-correlation coefficient (column 5) along with its error (column 6). Hardness ratio (HR) in column 7 along with position in the HID (Figure 2 (top)) and Lags (s) are shown in column 8. Full table is available in electronic version.}\\
\hline
\hline
\bf{OBSERVATION ID.}&\multicolumn{1}{|c|}{\textbf{START TIME}}&\multicolumn{1}{|c|}{\textbf{STOP TIME}}&{\bf EXPOSURE (S)}& {\bf CC}&{\bf CC ERROR}&{\bf HR}&{\bf LAG [ERROR] (S)}\\ \hline \hline
\endfirsthead

\multicolumn{8}{c}{{--continued from previous page}}\\

\hline
\hline
{\bf OBSERVATION ID.}&\multicolumn{1}{|c|}{\textbf{START TIME}}&\multicolumn{1}{|c|}{\textbf{STOP TIME}}&{\bf EXPOSURE (S)}& CCF&CCF ERROR&HR&LAG [ERROR] (S)\\
\hline
\endhead
\hline \multicolumn{6}{|r|}{{continued on next page}} \\ \hline
\endfoot

\hline
\hline
\endlastfoot
20056-02-01-00 & 02/02/1997 19:13:03  & 02/02/1197 19:58:03  & 2452 & 0.857 & 0.059 & 0.011 -- 0.023&-                                                   \\
20056-02-02-00       & 06/02/1997 22:46:13                                        & 07/02/1997 00:01:14                                       & 2511                                                             & -0.376                                                              & 0.126                                                 & 0.024-0.035 (II,III)    & -468.007 [-547.3, -389.0]                              \\
20053-03-02-00       & 07/02/1997 00:01:14                                        & 07/02/1997 03:20:13                                       & 5940                                                             & -                                                                   &                                                       & 0.022-0.033    & -                                                      \\
20053-03-02-01       & 08/02/1997 03:35:05                                        & 08/02/1997 05:52:13                                       & 32948                                                            & -                                                                   &                                                       & 0.022-0.033    & -                                                      \\
20053-03-02-02       & 08/02/1997 13:37:56                                        & 08/02/1997 14:12:13                                       & 1082                                                             & -                                                                   &                                                       & 0.025-0.034    & -                                                      \\
20056-02-03-00       & 10/02/1997 00:08:13                                        & 10/02/1997 00:58:13                                       & 2458                                                             & -0.431                                                              & 0.124                                                 & 0.053-0.070    & -                                                      \\
20056-02-04-00       & 13/02/1997 03:39:21                                        & 13/02/1997 05:14:13                                       & 2737                                                             & -0.385                                                              & 0.128                                                 & 0.027-0.039    & -                                                      \\
20056-02-05-00       & 15/02/1997 06:56:03                                        & 15/02/1997 07:36:13                                       & 2393                                                             & -0.299                                                              & 0.095                                                 & 0.014-0.023    & -                               \\
20056-02-06-00       & 18/02/1997 00:20:05                                        & 18/02/1997 01:00:13                                       & 2390                                                             & -0.402                                                              & 0.113                                                 & 0.034-0.047 (II)   & - 300.969 [-355.1,-248.4]                                                     \\
20056-02-07-00       & 23/02/1997 23:09:07                                        & 24/02/1997 00:12:13                                       & 2327                                                             & -0.309                                                              & 0.080                                                 & 0.012-0.019    & -                                                      \\
20056-02-08-00       & 27/02/1997 02:19:13                                        & 27/02/1997 03:34:13                                       & 2329                                                             & -                                                                   & -                                                     & 0.012-0.022    & -                                                      \\
20053-03-03-00 Sec A & 01/04/1997 19:13:46                                        & 02/04/1997 01:43:14                                       & 11471                                                            & 0.525                                                               & 0.091                                                 & 0.021-0.015    & -                                                      \\
“ Sec B              & \multicolumn{1}{>{\hspace{0pt}}p{0.138\linewidth}}{}       & \multicolumn{1}{>{\hspace{0pt}}p{0.138\linewidth}}{}      & \multicolumn{1}{>{\hspace{0pt}}p{0.087\linewidth}}{}             & 0.935                                                               & 0.028                                                 & “              & -                                                      \\
“ Sec C              & \multicolumn{1}{>{\hspace{0pt}}p{0.138\linewidth}}{}       & \multicolumn{1}{>{\hspace{0pt}}p{0.138\linewidth}}{}      & \multicolumn{1}{>{\hspace{0pt}}p{0.087\linewidth}}{}             & 0.827                                                               & 0.068                                                 & “              & -                                                      \\
“ Sec D              & \multicolumn{1}{>{\hspace{0pt}}p{0.138\linewidth}}{}       & \multicolumn{1}{>{\hspace{0pt}}p{0.138\linewidth}}{}      & \multicolumn{1}{>{\hspace{0pt}}p{0.087\linewidth}}{}             & 0.561                                                               & 0.012                                                 & “              & -                                                      \\
20053-03-03-01 Sec A & 02/04/1997 20:06:15                                        & 03/04/1997 00:12:14                                       & 8920                                                             & -0.599                                                              & 0.085                                                 & 0.060-0.037 (I,II)   & -207.366 [-241.0,-174.9]                               \\
“ Sec B              & \multicolumn{1}{>{\hspace{0pt}}p{0.138\linewidth}}{}       & \multicolumn{1}{>{\hspace{0pt}}p{0.138\linewidth}}{}      & \multicolumn{1}{>{\hspace{0pt}}p{0.087\linewidth}}{}             & -0.308                                                              & 0.103                                                 & “              & -668.750[-739.7,-595.6]                                \\
“ Sec C              & \multicolumn{1}{>{\hspace{0pt}}p{0.138\linewidth}}{}       & \multicolumn{1}{>{\hspace{0pt}}p{0.138\linewidth}}{}      & \multicolumn{1}{>{\hspace{0pt}}p{0.087\linewidth}}{}             & -0.478                                                              & 0.111                                                 & “              & -100 [-75, 125]                                        \\
20053-03-03-02 Sec A & 03/04/1997 21:46:09                                        & 04/04/1997 00:01:14                                       & 4967                                                             & -0.554                                                              & 0.093                                                 & 0.036-0.057    & -                                                      \\
“ Sec B              & \multicolumn{1}{>{\hspace{0pt}}p{0.138\linewidth}}{}       & \multicolumn{1}{>{\hspace{0pt}}p{0.138\linewidth}}{}      & \multicolumn{1}{>{\hspace{0pt}}p{0.087\linewidth}}{}             & -0.575                                                              & 0.136                                                 & “              & -                                                      \\
20053-03-03-03 Sec A & 04/04/1997 18:26:36                                        & 04/04/1997 23:26:13                                       & 9862                                                             & -                                                                   & -                                                     & 0.024-0.040 (II,III)   & ,-                                                     \\
“ Sec B              & \multicolumn{1}{>{\hspace{0pt}}p{0.138\linewidth}}{}       & \multicolumn{1}{>{\hspace{0pt}}p{0.138\linewidth}}{}      & \multicolumn{1}{>{\hspace{0pt}}p{0.087\linewidth}}{}             & 0.365                                                               & 0.084                                                 & “              & 1174.197 [1121.011,1226.001]                           \\
“ Sec C              & \multicolumn{1}{>{\hspace{0pt}}p{0.138\linewidth}}{}       & \multicolumn{1}{>{\hspace{0pt}}p{0.138\linewidth}}{}      & \multicolumn{1}{>{\hspace{0pt}}p{0.087\linewidth}}{}             & -                                                                   & -                                                     & “              & -                                                      \\
20053-03-01-00 Sec A & 27/07/1997 02:13:21                                        & 27/07/1997 08:02:14                                       & 11557                                                            & -                                                                   & -                                                     & 0.033-0.051    &                                                        \\
“ Sec B              & \multicolumn{1}{>{\hspace{0pt}}p{0.138\linewidth}}{}       & \multicolumn{1}{>{\hspace{0pt}}p{0.138\linewidth}}{}      & \multicolumn{1}{>{\hspace{0pt}}p{0.087\linewidth}}{}             & -0.545                                                              & 0.087                                                 & “              &                                                        \\
“ Sec C              & \multicolumn{1}{>{\hspace{0pt}}p{0.138\linewidth}}{}       & \multicolumn{1}{>{\hspace{0pt}}p{0.138\linewidth}}{}      & \multicolumn{1}{>{\hspace{0pt}}p{0.087\linewidth}}{}             & -                                                                   & -                                                     & “              &                                                        \\
“ Sec D              & \multicolumn{1}{>{\hspace{0pt}}p{0.138\linewidth}}{}       & \multicolumn{1}{>{\hspace{0pt}}p{0.138\linewidth}}{}      & \multicolumn{1}{>{\hspace{0pt}}p{0.087\linewidth}}{}             & -0.416                                                              & 0.132                                                 & “              &                                                        \\
20053-03-01-01 Sec A & 27/07/1997 16:41:07                                        & 27/07/1997 19:09:15                                       & 21736                                                            & -0.480                                                              & 0.101                                                 & 0.033-0.046 (II)   & -228.646 [-269.9,-188.3]                               \\
“ Sec B              & \multicolumn{1}{>{\hspace{0pt}}p{0.138\linewidth}}{}       & \multicolumn{1}{>{\hspace{0pt}}p{0.138\linewidth}}{}      & \multicolumn{1}{>{\hspace{0pt}}p{0.087\linewidth}}{}             & -0.341                                                                 & -0.110                                                     & “              & -                                                      \\
20053-03-01-02 Sec A & 27/07/1997 20:01:41                                        & 28/07/1997 00:33:14                                       & 9619                                                             & -0.757                                                              & 0.067                                                 & 0.034-0.049    & -                                                      \\
“ Sec B              & \multicolumn{1}{>{\hspace{0pt}}p{0.138\linewidth}}{}       & \multicolumn{1}{>{\hspace{0pt}}p{0.138\linewidth}}{}      & \multicolumn{1}{>{\hspace{0pt}}p{0.087\linewidth}}{}             & -0.571                                                              & 0.088                                                 & “              & -                                                      \\
“ Sec C              & \multicolumn{1}{>{\hspace{0pt}}p{0.138\linewidth}}{}       & \multicolumn{1}{>{\hspace{0pt}}p{0.138\linewidth}}{}      & \multicolumn{1}{>{\hspace{0pt}}p{0.087\linewidth}}{}             & -                                                                   & -                                                     & “              & -                                                      \\
\end{longtable*}
\endlongrotatetable




\clearpage

\begin{table}
\begin{minipage}[t]{\columnwidth}
\scriptsize
\caption{Best-fit spectral parameters for the spectra of ObsID 20053-03-03-01 and 70023-01-01-01. The letters A, B and C represent 
the sections in the light curve. The subscript BB represents the black body model, 
respectively. 
The unabsorbed flux in units of 10$^{-8}$ ergs cm$^{-2}$ s$^{-1}$ is calculated in the energy band 3--50 keV
and otherwise it is mentioned. Errors are quoted at a 90\% confidence level. Luminosity is in units of 10$^{38}$ erg s$^{-1}$, assuming the distance 13 kpc for GX 17+2.} 
\label{tab1}
\centering
\begin{tabular}{ccccccccc}
\hline
\hline
ObsID&&20053-03-03-01&&&70023-01-01-01&&\\
Parameters&A&B&C&A&B&C\\
\hline

\hline
&&&Black Body + nthcomp Model&&\\
\hline
$kT_{BB}$ (keV)& 3.01$\pm$0.15 & 2.73$\pm$0.04 &2.71$\pm$0.05 & 2.64$\pm$0.05& 2.71$\pm$0.02 &2.70$\pm$0.02 \\
$N_{BB}$& 0.08$\pm$0.03&0.08$\pm$0.01 &0.08$\pm$0.01& 0.06$\pm$0.02&0.08$\pm$0.008 &0.08$\pm$0.006\\
$\Gamma_{nthcomp}$&2.28$\pm$0.08 & 2.69$\pm$0.22& 2.67$\pm$0.22&2.34$\pm$0.24 & 2.73$\pm$0.16& 2.80$\pm$0.16\\
kT$_{soft}$& 0.70$\pm$0.12& 0.75$\pm$0.09&0.73$\pm$0.08& 1.08$\pm$0.30& 1.15$\pm$0.38&1.16$\pm$0.37\\
kT$_{e}$& 2.15$\pm$0.85& 5.85$\pm$2.22&5.82$\pm$2.20& 4.18$\pm$0.85& 6.66$\pm$1.52&7.59$\pm$2.01\\
N$_{nthcomp}$&2.83$\pm$0.16 & 2.92$\pm$0.10&2.90$\pm$0.10 &2.71$\pm$0.15 & 2.78$\pm$0.09&2.88$\pm$0.09 \\

BB flux&0.70&0.60 &0.58&0.49&0.64 &0.63\\
nthcomp flux&1.23& 1.30&1.31&1.54& 1.26&1.23\\
Total flux& 1.93 & 1.90 &1.89& 2.03 & 1.90 &1.86\\
$\chi^{2}$/dof&35/46 & 43/46&41/46&45/56 & 42/56&58/46\\
L$_{3-50 keV}$ &3.90 &3.84&3.82&4.10&3.84&3.76\\
R$_{BL}$ (km)&33.87&33.45&33.31&35.31&33.45&32.88\\
R$_{B}$ (km)&31.24&31.38&31.43&30.79&31.38&31.57\\

R$_{sp}$ (km)&33.99&33.46&33.29&35.75&33.46&32.75\\

\hline
&&&Model with cut-off power law&&\\
\hline
$kT_{BB}$ (keV)& 3.00$\pm$0.04 & 2.98$\pm$0.08 &3.00$\pm$0.08&2.98$\pm$0.18 & 2.77$\pm$0.05 &2.74$\pm$0.05 \\
$N_{BB}$& 0.08$\pm$0.008&0.061$\pm$0.005 &0.065$\pm$0.004& 0.05$\pm$0.009&0.06$\pm$0.008 &0.06$\pm$0.007\\
$\Gamma_{cutoffpl}$\footnote{Cut-off power-law index.}&0.93$\pm$0.16 & 1.43$\pm$0.11& 1.30$\pm$0.12& 1.13$\pm$0.24 & 1.76$\pm$0.17& 1.84$\pm$0.16\\
E$_{C}$\footnote{Cut-off energy.}(keV)& 3.40$\pm$0.55& 5.94$\pm$0.65&5.64$\pm$0.73&5.31$\pm$1.01& 8.32$\pm$1.15&8.93$\pm$1.45\\
N$_{cutoffpl}$\footnote{Normalization.}&5.06$\pm$0.42 & 5.75$\pm$0.44&5.67$\pm$0.43&4.36$\pm$0.75 & 7.23$\pm$1.32&7.71$\pm$1.42\\

BB flux&6.4&4.9&5.0&4.15&5.11 &4.92\\
Cutoffpl flux&12.9&14.2&14.0&16.32& 13.91&13.62\\
Total flux& 19.6&19.4&19.4&20.72 &19.42 &19.14\\
L$_{3-50 keV}$&3.96&3.92& 3.92&4.17& 3.91&3.84 \\
$\chi^{2}$/dof&21/46&24/46&24/46&41/56 & 48/56&61/56\\

\hline
\end{tabular}
\end{minipage}
\end{table}

\clearpage

\begin{table}
\begin{minipage}[t]{\columnwidth}
\scriptsize
\caption{Same as for Table 2 but for ObsID 80022-01-05-00 in energy band 3--40 keV. Flux is in units of 10$^{-9}$ ergs cm$^{-2}$ s$^{-1}$. }
\label{tab1}
\centering
\begin{tabular}{ccccccccc}
\hline
\hline
Parameters&A&B&C&D&E\\
\hline

\hline

$kT_{BB}$ (keV)&1.02$\pm$0.03& 0.98$\pm$0.03 &1.00$\pm$0.03&0.93$\pm$0.03 &0.91$\pm$0.03 \\
$N_{BB}$& 0.02$\pm$0.008&0.03$\pm$0.008 &0.03$\pm$0.003& 0.03$\pm$0.008&0.05$\pm$0.008\\
$\Gamma_{nthcomp}$\footnote{asymptotic power-law of thermal Comptonization model}&2.03$\pm$0.05 & 1.95$\pm$0.05& 1.98$\pm$0.05&1.91$\pm$0.05&1.81$\pm$0.05\\
kT$_{e}$& 3.20$\pm$0.95& 3.10$\pm$0.96&3.14$\pm$0.95&3.14$\pm$0.95 &3.00$\pm$0.93\\
N$_{nthcomp}$&2.36$\pm$0.20 & 2.18$\pm$0.21& 2.22$\pm$0.21&2.11$\pm$0.18 &1.79$\pm$0.19 \\
BB flux& 3.67&3.92&3.83&3.78&2.36\\

nthcomp flux& 14.69 & 14.35&14.30&14.28&15.91\\
Power-law flux&-- &--&--&--&1.10 (fixed)\\
Norm$_{PL}$&-- &--&--&--&\\
Total flux& 18.62 & 18.63 &18.43& 18.41&18.65\\
L$_{3-40 keV}$ & 3.76& 3.76&3.73&3.72&3.77 \\
R$_{BL}$ km&32.91&32.92&32.64&32.62&32.95\\
R$_{B}$ km &31.56&31.56&31.65&31.66&31.54\\
R$_{sp}$ km&32.79&32.81&32.45&32.42&32.84\\
$\chi^{2}$/dof&39/53 & 34/53&50/53& 38/53&36/53\\

\hline
\hline
&&Model with Cut-off Power-law&\\
\hline
\hline
$kT_{BB}$ (keV)& 0.95$\pm$0.02 & 0.85$\pm$0.02 &0.86$\pm$0.02&0.84$\pm$0.02 &0.83$\pm$0.02 \\
$N_{BB}$& 0.07$\pm$0.007&0.08$\pm$0.008 &0.08$\pm$0.008&0.09$\pm$0.006&0.08$\pm$0.006\\
$\Gamma_{cutoffpl}$&0.16$\pm$0.10 & 0.20$\pm$0.11& 0.27$\pm$0.10&0.15$\pm$0.12 &0.25$\pm$0.10\\
E$_{C}$& 3.97$\pm$0.12& 3.97$\pm$0.13&4.02$\pm$0.12&4.06$\pm$0.13 &4.17$\pm$0.13\\
N$_{cutoffpl}$&0.93$\pm$0.20 & 1.05$\pm$0.22&1.15$\pm$0.30&0.86$\pm$0.13 &1.01$\pm$0.20\\

BB flux&3.58&3.55 &3.35&3.48 &3.16\\
Cutoffpl flux&14.24& 14.73&14.85&14.47 &14.69\\
Total flux& 18.17 &18.62 &18.43& 18.37&18.47\\
L$_{3-40 keV}$& 3.67& 3.76&3.72&3.71&3.73 \\
$\chi^{2}$/dof&62/53 & 32/53&52/53&43/53&61/53\\

\hline
\hline
\end{tabular}
\end{minipage}
\end{table}

\clearpage

\clearpage

\begin{table}
\begin{minipage}[t]{\columnwidth}
\tiny
\caption{Best-fit spectral parameters for the spectra of ObsID 20053-03-03-01 and 70023-01-01-01. The letters A, B and C represent 
the sections in the light curve. The unabsorbed flux in units of 10$^{-8}$ ergs cm$^{-2}$ s$^{-1}$ is calculated in the energy band 3--25 keV
and otherwise it is mentioned.} 
\label{tab1}
\centering
\begin{tabular}{cccccccccccc}
\hline
\hline
ObsID&&20053-03-03-01&&&70023-01-01-01&&\\
\hline
Parameters&A&B&C&A&B&C\\
\hline

\hline

$kT_{diskbb}$ (keV)\footnote{Temperature of the Diskbb model.}& 1.25$\pm$0.10& 0.92$\pm$0.10 &0.92$\pm$0.10 &1.47$\pm$0.09& 1.09$\pm$0.16 &1.04$\pm$0.06 \\
$N_{diskbb}$\footnote{Normalization of the Diskbb model.}& 127$\pm$30&334$\pm$154 &323$\pm$162& 32$\pm$7&129$\pm$70 &145$\pm$89\\
$N_{diskbb}*$\footnote{Normalization of the Diskbb model after freezing the nthcomp parameters.}&127$\pm$14&334$\pm$112&323$\pm$115&32$\pm$6&130$\pm$44&145$\pm$58\\
$\Gamma_{nthcomp}$\footnote{Asymptotic power-law of the thermal Comptonization model.}&1.83$\pm$0.08 & 1.86$\pm$0.04& 1.88$\pm$0.14&1.82$\pm$0.07 & 1.83$\pm$0.06& 1.86$\pm$0.06\\
kT$_{e}$\footnote{Electron temperature.}(keV)& 3.22$\pm$0.90& 3.41$\pm$0.86&3.41$\pm$0.86& 3.27$\pm$0.92& 3.39$\pm$0.80&3.42$\pm$0.82\\
N$_{nthcomp}$\footnote{Normalization.}&1.80$\pm$0.32 & 1.95$\pm$0.15&1.98$\pm$0.15&1.94$\pm$0.02 & 1.80$\pm$0.24&1.85$\pm$0.22 \\

Diskbb flux&0.24&0.12 &0.10&1.39&0.12 &0.10\\
nthcomp flux&1.69& 1.79&1.79&1.88& 1.73&1.71\\
Total flux& 1.93 & 1.91 &1.89& 3.27 & 1.85 &1.81\\
L$_{3-25 keV}$ & 3.90& 3.86&3.82& 6.61& 3.74&3.65  \\

R$_{in}$ (km) cos(\it{i}=25$^{o}$) & 18.46$\pm$ 1.02& 29.94$\pm$5.09&29.45$\pm$5.33&9.26$\pm$1.00& 18.67 $\pm$ 3.20&19.72$\pm$4.03 \\
cos(\it{i}=35$^{o}$) & 19.41$\pm$2.20& 31.50$\pm$5.90&30.97$\pm$6.19&9.74$\pm$1.02& 19.64$\pm$3.73 &20.75$\pm$4.60 \\
$\chi^{2}$/dof&35/46 & 42/46&42/46&29.38/41 & 24.88/41&41.76/41\\
\hline

Delay (s)&--207$\pm$34& --668$\pm$72& -100$\pm$25&114$\pm$26 &--&-168$\pm$75&\\
$\nu$ (Hz)&32.32$\pm$1.26&24.73$\pm$0.71&24.05$\pm$0.77&26.64$\pm$1.14&24.40$\pm$1.02&25.39$\pm$1.05\\
R$_{in}$ (km) \footnote{Using equation 1.}&21.01$\pm$0.28&23.07$\pm$0.23&23.28$\pm$0.26&22.47$\pm$0.28&23.17$\pm$0.19&22.85$\pm$0.13\\
H$_{corona}$ (km) $\beta$=0.05 \& 0.1\footnote{Substituting R$_{in}$ value obtained from equation 1.}&17.67 \& 34.14 &55.99 \& 111.99& 7.35 \& 14.71&8.87 \& 17.78 & --& 12.95 \& 25.90\\


H$_{corona}$ (km) $\beta$=0.05 \& 0.1\footnote{Substituting R$_{in}$ value diskbb normalization.} &17.81\& 35.63 &49.07 \& 98.14& 6.08 \& 12.16&--\&  & --& 13.80 \& 27.60\\
H$_{corona}$ (km) $\beta$=0.05 \& 0.1\footnote{If disk is at the last stable orbit i.e. 12 km.} &25.42\& 50.84 &84.87 \& 169.74& 12.19  \& 24.39 & 14.87\& 28.37 & --& 20.59  \& 41.18\\


\hline


\hline
\end{tabular}
\end{minipage}
\end{table}

\clearpage
\begin{table}
\caption{Results of spectral fit for section A's initial (A) and final (B) parts of ObsID 70023-01-01-01 (see Text).}
\label{Table 2}
\small
\centering
\begin{tabular}{ccccc}
\hline
\hline
Parameters & A & B  \\
\hline
$\Gamma$$_{nthcomp}$& 1.87$\pm$0.02 &1.98$\pm$0.02\\
kT$_{e} (keV) $&3.29$\pm$0.85&3.48$\pm$0.87\\
kT$_{soft}$&0.97$\pm$0.13&0.96$\pm$0.12\\
N$_{nthcomp}$&2.42$\pm$0.20 & 2.62$\pm$0.26 \\
$\chi^{2}$/dof& 37/42&44/42\\
\hline
\hline
\end{tabular}
\end{table}


\begin{table}
\begin{minipage}[t]{\columnwidth}
\scriptsize
\caption{Spectral parameters of NuSTAR FPMA and FPMB combined spectra in energy band 3--30 keV. Flux is in units of 10$^{-8}$ ergs cm$^{-2}$ s$^{-1}$. Luminosity is calculated in units of 10$^{38}$ erg s$^{-1}$, assuming the distance 13 kpc for GX 17+2. }
\label{tab1}
\centering
\begin{tabular}{cc|cc}
\hline
\hline
Parameters\\
\hline
\hline
	& Black Body+nthcomp && Diskbb+nthcomp\\
\hline
$kT_{BB}$ (keV)&1.09$\pm$0.04 & $kT_{diskbb}$ (keV)&1.39$\pm$0.06 \\
$N_{BB}$& 0.03$\pm$0.01&$N_{diskbb}$&73$\pm$8\\
R$_{in}$ cos (i=25$^\circ$)& &&14.01$\pm$1.54\\
R$_{in}$ cos (i=35$^\circ$)& &&14.73$\pm$1.62\\
$\Gamma_{nthcomp}$&1.98$\pm$0.08 & $\Gamma_{nthcomp}$&2.00$\pm$0.07\\
kT$_{e}$& 3.11$\pm$1.01& kT$_{e}$& 3.15$\pm$0.95\\
kT$_{soft}$&$=kT_{BB}$ && = $kT_{diskbb}$\\ 
N$_{nthcomp}$&1.69$\pm$0.25 & N$_{nthcomp}$&1.21$\pm$0.13 \\
BB flux& 0.15& Diskbb flux&0.25\\

nthcomp flux& 1.37&nthcomp flux& 1.28\\
Total flux& 1.54 &Total flux&1.54\\
L$_{3-30 keV}$ & 3.11&L$_{3-30 keV}$& 3.11\\
$\chi^{2}$/dof&573/542 &$\chi^{2}$/dof& 575/542\\
\hline
\hline

\end{tabular}
\end{minipage}
\end{table}


\label{lastpage}

\end{document}